\documentclass[pre,twocolumn,twoside,showpacs,superscriptaddress,floatfix]{revtex4-1}
\usepackage{graphics,amssymb,amsmath,epsf,color,hyperref}
\epsfclipon
\usepackage{epsfig,bm,epsf,graphics}
\usepackage{graphicx}% Include figure files
\usepackage{dcolumn}% Align table columns on decimal point
\usepackage{bm}% bold math
\usepackage{braket}% braket
\usepackage{color}
\usepackage{soul}
\usepackage{xcolor}
\newcommand{\rev}{\textcolor{black}}
\newcommand{\revi}{\textcolor{black}}
\begin{document}

%\title{Restoration of Time Series Data Using Kinetic Ising Model}
\title{Inference of stochastic time series with missing data}

\author{Sangwon Lee}
\affiliation{Department of Physics and Astronomy, Seoul National University, Seoul 08826, Korea}

\author{Vipul Periwal}
\email[Corresponding author: ]{vipulp@mail.nih.gov}
\affiliation{Laboratory of Biological Modeling, National Institute of Diabetes and Digestive and Kidney Diseases, National Institutes of Health, Bethesda, Maryland 20892, USA}

\author{Junghyo Jo}
\email[Corresponding author: ]{jojunghyo@snu.ac.kr}
\affiliation{Department of Physics Education and Center for Theoretical Physics and Artificial Intelligence Institute, Seoul National University, Seoul 08826, Korea}
\affiliation{School of Computational Sciences, Korea Institute for Advanced Study, Seoul 02455, Korea}

\date{\today}

\begin{abstract}
Inferring dynamics from time series is an important objective in data analysis.
In particular, it is challenging to infer stochastic dynamics given incomplete data.
We propose an expectation maximization (EM) algorithm that iterates between alternating two steps: E-step restores missing data points, while M-step infers an underlying network model from the restored data.
Using synthetic data of a kinetic Ising model, we confirm that the algorithm works for restoring missing data points as well as inferring the underlying model.
At the initial iteration of the EM algorithm, the model inference shows better model-data consistency with observed data points than with missing data points.
As we keep iterating, however, missing data points show better model-data consistency. 
We find that demanding equal consistency of observed and missing data points provides an effective stopping criterion for the iteration to prevent going beyond the most accurate model inference.
\rev{Using the EM algorithm and the stopping criterion together}, we infer missing data points from a time-series data of real neuronal activities.
\rev{Our method reproduces collective properties of neuronal activities such as correlations and firing statistics even when 70\% of data points are masked as missing points}.%, which have previously never been optimized to fit.
\end{abstract}

%In this paper, we propose a methodology to perform inference of kinetic Ising model when a fraction of data have been lost. This method is based on the recent work by Hoang \textit{et al.} in which the missing observations and model parameters are optimized alternately by the expectation-maximization algorithm. To avoid overfitting, we set a stopping criterion of our iterative algorithm using quality-of-fit measures of both visible and hidden data points. We tested our method on synthetic data of the Sherrington-Kirkpatrick model and found that the algorithm correctly infers both model parameters and pair correlations without bias. Turning to real data, we applied our algorithm to neural spike train datasets and found that it successfully recovers collective properties of the data. Overall, we expect that our methodology can be widely used in many applications.

%\pacs{05.45.Xt, 05.45.-a, 05.70.Fh}
%05.45.Xt: coupled oscillators, synchronization, nonlinear dynamics
%05.45.-a: dynamical systems, nonlinear dynamics
%05.70.Fh: phase transition in statistical mechanics and thermodynamics
 
\maketitle

%\tableofcontents

\section{Introduction}

System identification is one of the most important tasks in data science~\cite{Brunton2019}. To be specific, suppose we have observations $\{ \vec{\sigma}(t)\}$ of a sample $\vec{\sigma} = (\sigma_1, \sigma_2, \cdots, \sigma_N)$, where $t$ can denote a time index for time series data, or a sample index for independent data. Considering stochastic systems, statistical mechanics has been adopted to construct the least structured probabilistic model of $P(\vec{\sigma})$ from observations. In particular, Ising-like models of $P(\vec{\sigma}) \propto \exp \big(\sum_{i,j} W_{ij} \sigma_i \sigma_j \big)$ incorporate the pair-wise interactions between variables. Its applicability ranges from neuroscience and biology~\cite{Schneidman2006Apr,Shlens2006Aug,Lapedes2012Jul,Morcos2011Dec,Mora2010Mar,Lezon2006Dec} to economics and sociology~\cite{Bury2013Mar,Shemesh2013Sep,Lee2015Jul}.
Such pairwise interactions are generally sufficient to explain complex higher-order patterns in many cases~\cite{Schneidman2006Apr,Figliuzzi2018Apr}.
%and the efficacy and limitations of these models have also been extensively studied~\cite{Roudi2009May}.
To consider time-dependent data arising from kinetic interactions, another type of Ising model has been proposed~\cite{Derrida1987Jul}.
Unlike the equilibrium model of $P(\vec{\sigma})$, the kinetic Ising model has a probabilistic relation between $\sigma_i(t+1)$ and $\sigma_j(t)$ with the conditional probability of $P(\sigma_i(t+1)|\vec{\sigma}(t)) \propto \exp\big(\sum_{j} W_{ij} \sigma_i(t+1) \sigma_j(t)\big)$.
%There is a variant of an Ising model with nonequilibrium dynamics, called \textit{kinetic Ising model}~\cite{Derrida1987Jul}. Unlike other Ising-like models describing a probability distribution of each observed pattern, this model deals with the evolution of the sequence with time and 
The kinetic model has been used to reconstruct neural networks from temporal neuronal activities~\cite{Roudi2011Jan,Zeng2013May}.

Both equilibrium and kinetic (or non-equilibrium) models have network parameters $W_{ij}.$
In the equilibrium model, it is symmetric ($W_{ij} = W_{ji}$) to represent undirected correlation between $\sigma_i$ and $\sigma_j$. However, in the kinetic model, $W_{ij}$ is not necessarily symmetric ($W_{ij} \neq W_{ji}$) as it represents directed causality from $\sigma_j(t)$ to $\sigma_i(t+1)$.
Due to the wide applications of these models, it is an important inverse problem to infer $W_{ij}$ from observations $\{ \vec{\sigma}(t)\}$.
As concretely established in the equilibrium model~\cite{Nguyen2017Jul}, the kinetic model also has many inference methods including various mean-field approximations~\cite{Roudi2011Jan,Mezard2011Jul}, maximum likelihood estimation~\cite{Roudi2011Jan,Zeng2013May}, and the recent expectation-reflection \rev{(ER)} method~\cite{Hoang2019Feb}. % These algorithms might be helpful to reconstruct networks of interactions that appear in many scientific areas.

In real-world problems, it is common that only a part of a network is observable.
For example, it is impossible to observe every neuron in the brain.
Therefore, one should consider not only observed visible units but also unobserved hidden units. Much effort has been devoted to infer both hidden variables and the network parameters~\cite{Dunn2013Feb,Tyrcha2014,Bachschmid-Romano2014Jun,Battistin2015May,Hoang2019Apr}.
%There have also been many efforts to apply the kinetic Ising model when only a part of a network is observable~\cite{Dunn2013Feb,Tyrcha2014,Bachschmid-Romano2014Jun,Battistin2015May,Hoang2019Apr}. This situation is common in neuroscience since we cannot observe all existing neurons. In this case, we should consider unobserved hidden units along with observed visible units, and various methods have been designed to infer both these hidden variables and the model parameters. 
These methods basically rely on the expectation-maximization (EM) algorithm~\cite{Dempster1977Sep}.
The EM method is composed of two iterative steps: 
E-step predicts hidden variables by using mean-field approximations~\cite{Dunn2013Feb,Tyrcha2014} or replica-based approaches~\cite{Bachschmid-Romano2014Jun,Battistin2015May}, or a likelihood-based method~\cite{Hoang2019Apr}, 
and M-step optimizes the network parameters from the reconstrcted data.

In addition to the issue of hidden units, another practical issue is that even visible units are not always observable throughout an experiment.
%Recently, Campajola \textit{et al.} suggested a somewhat different scenario ~\cite{Campajola2019Jun}. Instead of splitting a network into observed and unobserved regions, all nodes in a network are measured throughout an experiment but not permanently. 
At each time point, some units become accessible to observers and the others become inaccessible. 
For example, a large-scale neural network can be partially scannable in neuroscience experiments~\cite{Soudry2015Oct}.
This scenario is also common in finance and social science.
For a trading network, trade records are available only when traders are active~\cite{Campajola2020Nov}. 
%, and a similar approach to the above problem can be applied. For instance, the authors of~\cite{Campajola2019Jun} used their own methods to infer trader networks from trade records whose agents can be either active or inactive~\cite{Campajola2020Nov}. Other examples involve the reconstruction of a large-scale neural network~\cite{Soudry2015Oct}. Rather than observing the entire population which is infeasible, Soudry \textit{et al.} proposed a ``shotgun'' approach to measure multiple subsets of neurons successively.
Although some pioneering work exists, research on the kinetic Ising model with missing or partially masked data is still in its infancy.
%and more studies need to be done. For example, the methodology 
Campajola {\it et al.}~\cite{Campajola2019Jun} addressed this issue inspired by the mean-field approach of Dunn {\it et al.}~\cite{Dunn2013Feb} that was originally developed for the problem of hidden units.
The mean-field approach imposes an \textit{a priori} assumption that interactions are weak and dense~\cite{Tyrcha2014}. Thus it is imperative to develop an approach free from such constraints. 
Of course, any such method has to be tested against real-world problems beyond the proof-of-concept with synthetic data.
%For example, it is of particular interest to see whether the model inference can reproduce collective behaviors of neurons from neural activity data ~\cite{Tkacik2014Jan,Chen2019May}.
%, so other approaches free from these constraints should be developed. Also, an important question is to what extent our statistical physics model can mimic underlying structures of data like collective behavior~\cite{Tkacik2014Jan,Chen2019May} or correlations between nodes. Application of the model not only to synthetic data but also to real-world data should also be thoroughly tested.

Here we develop a general method to infer the parameters of a kinetic Ising model from data with sporadic missing values of any measured variable, extending our recent study on hidden units~\cite{Hoang2019Apr}.
The algorithm uses the EM method: the E-step restores missing time points stochastically using a likelihood ratio, while the M-step infers network parameters from observed and restored data. 
\rev{Approximating the E-step with a stochastic realization is called the stochastic approximation EM (SAEM)~\cite{Delyon1999}.
By repeating the alternation between E- and M-steps, the algorithm infers network parameters.
However, too much iteration can overfit the model, preventing accurate inference of model parameters.
It is therefore crucial to find a criterion for stopping the EM iterations at the right time.}
%Alternation of these two steps infers network coupling parameters as we iterate but excessive iteration finally ends up with overestimation of the parameters. It is therefore crucial to find a criterion for stopping EM iterations that can be applied to real-world data. 
\rev{We find an effective stopping criterion for such an optimal inference.
It is based on the intuition that once missing data is properly restored, there should not be any distinction between observed and restored data.
Therefore we stop the iteration when the stochastic model shows equal uncertainty for the prediction of observed and missing data points.}
Using the EM method with such an optimal number of iterations, we demonstrate that our method successfully infers interactions from synthetic data of the kinetic Ising model. We then apply the method to infer the kinetics of a neuronal network from real neuronal activity data~\cite{Tkacik2014Jan,Chen2019May}, and confirm that the inference reproduces collective behaviors such as time correlation and firing statistics of neuronal activities.

This paper is organized as follows.
In Section~\ref{method}, we describe our EM-based method and the stopping criterion for optimal iterations. In Section~\ref{results}, we validate this method with simulated data using the kinetic Ising model.
%and examine the inference performance \rev{under various conditions}. 
We then apply the method to infer dynamics from a recording of neuronal activity, and compare the inference performance with equilibrium and simple imputation models. Finally, we summarize and discuss our findings in Section~\ref{discussion}.
To keep the paper self-contained, \rev{we briefly introduce the mean-field method developed by Campajola \textit{et al.}~\cite{Campajola2019Jun} as a benchmark of our method in Appendix~\ref{appendix_campajola}. We also consider various conditions to validate our method in Appendix~\ref{appendix_experiment}. Furthermore, we explain the equilibrium model in Appendix~\ref{appendix_eq},} and describe additional experimental data on neuronal activities in Appendix~\ref{appendix_neuron}. 

%In this paper, we propose a novel method inspired by Hoang \textit{et al.}~\cite{Hoang2019Apr} to infer the kinetic Ising model with partially missing data, which does not need additional requirements such as the weak coupling assumption. 
%In the E-step, missing observations are stochastically assigned using a likelihood ratio, following by the M-step where model parameters are optimized from the recovered data. Alternating these two steps, we observed evidence of overfitting after few iterations. To avoid overfitting, we compared quality-of-fit measures of visible and hidden data points and stopped the algorithm when these two functions become close to each other. We checked that this algorithm showed a reasonable performance with synthetic kinetic Ising data with various hyperparameters such as the ratio of missing observations, a coupling strength, and a sample size. Then we applied our method to actual neural activity data, which do not follow the strict kinetic Ising model, to validate it. Additionally, we provide a similar data restoration algorithm with an equilibrium Ising model for completeness and compare the performance of the two algorithms.

\section{Method}
\label{method}

We consider stochastic dynamics of the kinetic Ising model with $N$ binary variables.
The $i$-th spin $\sigma_i(t+1)$ at time $t+1$ is stochastically determined by the conditional probability:
\begin{equation}
\label{eq:prob}
P(\sigma_i(t+1)=\pm1|\vec{\sigma}(t), \theta) = \frac{\exp[\pm H_i(t)]}{\exp[H_i(t)]+\exp[-H_i(t)]},
\end{equation}
where the local field $H_i(t) = \sum_j W_{ij} \sigma_j(t) + b_i$ integrates the influence of connected spins as well as external bias $b_i$. \rev{We denote the model parameters as $\theta = (W_{ij}, b_i)$.}
Here we choose a synchronous update for simplicity; refer to~\citep{Zeng2011Apr,Zeng2013May} for an asynchronous update.
Given binary time series data \rev{$\{ \vec{\sigma}(t) \} _{t=0}^L$ of length $L+1$}, various methods exist to infer $\theta$ in Eq.~(\ref{eq:prob})~\cite{Roudi2011Jan,Nguyen2017Jul,Mezard2011Jul,Hoang2019Feb}.
In particular, the \rev{ER} method provides faster inference than the standard maximum likelihood estimation, since the iterative algorithm is based on a multiplicative and parallelizable  parameter update~\cite{Hoang2019Apr}.
In this study, however, we used standard logistic regression included in the Python package, \textit{scikit-learn}~\cite{scikit-learn}, because logistic regression shows similar performance with the \rev{ER} method with high accuracy and fast computation.
Note that Eq.~(\ref{eq:prob}) implies the logistic function:
\begin{equation}
\label{eq:logistic}
P(\sigma_i(t+1)=1|\vec{\sigma}(t), \theta) = \frac{1}{1 + \exp[-2 \sum_j W_{ij} \sigma_j(t) -2 b_i]}.
\end{equation}
\noindent \rev{In the absence of regularization, the logistic regression is equivalent to the maximization of a total likelihood of data,
\begin{equation}
\label{eq:likelihood}
\mathcal{L}_{\mathrm{tot}}(\theta) = \prod _{i=1}^N \prod_{t=0}^{L-1} P(\sigma_i (t+1) | \vec{\sigma} (t), \theta).
\end{equation}
}
%First, we provide a basic setting of the kinetic Ising model with {\em N} number of spins. The state of spins at time {\em t} is represented by $\vec{\sigma}(t) = (\sigma_1(t), \sigma_2(t), ..., \sigma_N(t))$. All spins are binary, \textit{i.e.} $\sigma_i(t) = \pm 1$. For each spin $i$ and time $t$, a local field $H_{i}(t)$ is given by $H_{i}(t) = \sum_{j} W_{ij}\sigma_j(t)$, where $W_{ij}$ denotes a coupling parameter (we can apply an additional external field $h_i$, but it does not change the whole picture). The state of spins at time $t+1$ is stochastically determined from $H_i (t)$ by
%\begin{equation}
%\label{eq:prob}
%P(\sigma_i(t+1)=\pm1|\vec{\sigma}(t)) = \frac{\exp[\pm H_i(t)]}{\exp[H_i(t)]+\exp[-H_i(t)]}.
%\end{equation}

Now suppose that some of the data points are missing. Let $\mathcal{M}$ denote the set of missing data points. 
\rev{We explicitly distinguish between missing and observed data points by denoting $\sigma^m_i(t)$ for $(i,t) \in \mathcal{M}$ and $\sigma^o_i(t)$ for $(i,t) \notin \mathcal{M}$. However, when the distinction is not necessary, we omit the superscripts $m$ and $o$.}
To recover these missing values, we first assign random binary values for \rev{$\sigma^m_i(t)$} at every $(i,t) \in \mathcal{M}$.
Then, from the observed and randomly-assigned values of $\{ \vec{\sigma}(t) \} _{t=0}^L$, we can infer an initial value of \rev{$\theta$} using logistic regression in Eq.~(\ref{eq:logistic}), \rev{or maximizing $\mathcal{L}_{\mathrm{tot}}(\theta)$ in Eq.~(\ref{eq:likelihood})}.
%Next, we define a likelihood function for $\sigma _i (t) = \pm 1$ ($t \ne 1$ or $L$) as
\rev{The next step is to update the missing data points. Previous EM-based algorithms used a mean-field approach to approximate missing data points as mean values~\cite{Dunn2013Feb,Campajola2019Jun}. 
For the E-step, however, we stochastically assign missing data points following SAEM.
The likelihood of missing values $\sigma^m_i (t)= \pm 1$ can be derived from the total likelihood $\mathcal{L}_{\mathrm{tot}}$ as follows,}
\begin{equation}
\label{eq:update}
\mathcal{L}_{i,t}^{\pm} \equiv P(\sigma^m_i (t) = \pm 1 | \vec{\sigma} (t-1),\theta) \prod _{j=1} ^N P(\sigma_j (t+1) | F_i^{\pm} (\vec{\sigma} (t),\theta),
\end{equation}
where $F_i^{\pm} (\vec{\sigma} (t)) = (\sigma_1 (t), \cdots, \sigma^m_{i} (t) = \pm 1, \cdots, \sigma_N (t))$ \rev{for $(i,t) \in \mathcal{M}$}.
%\rev{Eq.~(\ref{eq:update}) was first proposed by Hoang \textit{et al.} \cite{Hoang2019Apr} to restore hidden spins from a network of visible spins.}
$\mathcal{L}_{i,t}^{\pm}$ is a product of the likelihoods determined by the one-step backward state of $\vec{\sigma} (t-1)$ and the one-step forward state $\vec{\sigma} (t+1)$. 
The likelihoods for \rev{$t = 0$} and $L$ involve only forward and backward states, respectively:
\rev{$\mathcal{L}_{i,0}^{\pm} \equiv \prod _{j=1} ^N P(\sigma_j (1) | F_i^{\pm} (\vec{\sigma} (0)), \theta)$ and $\mathcal{L}_{i,L}^{\pm} \equiv P(\sigma^m_i (L) = \pm 1 | \vec{\sigma} (L-1),\theta)$.}
%At $t=1$, only a likelihood for the forward we only think about a likelihood from the future spins. On the other hand, when $t=L$, we simply define the likelihood for $\sigma_i (L)$ as $P(\sigma_i (L) | \vec{\sigma} (L-1))$. 
Using these likelihood values, we stochastically re-assign $\pm 1$ to \rev{$\sigma^m_i (t)$} with a probability of $\mathcal{L}_{i,t}^{\pm}/(\mathcal{L}_{i,t}^{+}+\mathcal{L}_{i,t}^{-})$ for every missing data point of $(i,t) \in \mathcal{M}$ \rev{with random order}. 
\rev{Alternatively, we have updated the missing data points with a Metropolis-like manner~\cite{Hastings1970Apr}, and confirmed no noticeable changes (data not shown).}
Also note that updating one missing point is affected by other missing points that may or may not have been updated in that step.

After updating all missing data points, we optimize \rev{$\theta$ from the observed and restored data $\sigma_i(t)$ by maximizing $\mathcal{L}_{\text{tot}}(\theta)$ in Eq.~(\ref{eq:likelihood}) (\rev{M}-step).}
%In~\cite{Hoang2019Apr}, the authors used the free-energy minimization method~\cite{Hoang2019Feb} to find $W_{ij}$ since it was an order of magnitude faster than the maximum likelihood estimation while giving more reliable results. Alternatively, we used a logistic regression function included in a Python package \textit{scikit-learn}~\cite{scikit-learn}. Since Eq.~\ref{eq:prob} has a form of logistic regression, this function can be readily used to infer the kinetic Ising model with high accuracy and low computation time. Nevertheless, we note here that using the free-energy minimization method in this work also produced similar results.
Repeating these E- and M-steps, \rev{$\theta$ is expected to converge to the true parameter values. However, excess iteration ends up with worse inference for $\theta$}, especially when a larger fraction of data is missing, a well-known issue in the latent variable field~\cite{Rubin2002}. %{\color{red}{You can include a reference to Rubin's paper here, see Julie Josse's website for the reference.}}
%This is different from typical overfitting issues because the model complexity does not play any role in the fitting process. The $W_{ij}$ just keeps increasing and no regularization techniques are applicable. 
\rev{How to stop the EM iteration at a right time?}
%Real-world problems do not come with known true $W_{ij},$ and the only way to avoid overestimation is to stop the iteration at the proper time without prior knowledge of $W_{ij}$.
%How can we figure out the optimal number of iterations maximizing the goodness of fit of the inferred $W_{ij}$?}

%\rev{When it comes to missing data imputation, it is essential to demand an equal degree of uncertainty of both observed and missing data points. Several metrics like cross-entropy can be used to quantify the degree of uncertainty. However, in this study,}
\rev{To address this issue, we consider model-data consistency.
It is not entirely straightforward to check consistency in stochastic models.
In the kinetic Ising model, the future state $\sigma_i(t+1)$ probabilistically depends on the current state $\vec{\sigma}(t)$.
Therefore we consider the discrepancy between $\sigma_i(t+1)$ and its expectation value,}
\begin{align}
\label{eq:expectation_sigma}
\mathbb{E} [\sigma_i (t+1)] = & \sigma_i(t+1) P(\sigma_i(t+1)| \vec{\sigma}(t),\theta) \nonumber \\
& - \sigma_i(t+1) \big[ 1- P(\sigma_i(t+1)| \vec{\sigma}(t), \theta) \big] \nonumber \\
= & \sigma_i(t+1) \big[ 2P(\sigma_i(t+1)| \vec{\sigma}(t),\theta) - 1\big],
\end{align}
as a quality-of-fit measure. 
The expectation value is simply obtained as $\mathbb{E} [\sigma_i(t+1)] = \tanh{H_i (t)}$ for the kinetic Ising model following the conditional probability of Eq.~(\ref{eq:prob}).
%\rev{This was first used in \cite{Hoang2019Feb} to set a stopping criterion for the ER algorithm.} 
\rev{Here the mean discrepancy for every data point can be quantified as
\begin{align}
\label{eq:D}
D & = \rev{\frac{1}{NL}} \sum_{i=1}^N \sum_{t=0}^{L-1} \big[\sigma_i (t+1) - \mathbb{E} [\sigma_i (t+1)] \big]^2 \nonumber \\
& = \rev{\frac{1}{NL}} \sum_{i=1}^N \sum_{t=0}^{L-1} \sigma_i^2(t+1) \big[ 2 - 2P(\sigma_i(t+1)| \vec{\sigma}(t),\theta) \big]^2 \nonumber \\
& = \rev{\frac{4}{NL}} \sum_{i=1}^N \sum_{t=0}^{L-1} \big[ 1 - P(\sigma_i(t+1)| \vec{\sigma}(t),\theta) \big]^2.
\end{align}
}For the derivation of the second line in Eq.~(\ref{eq:D}), we used Eq.~(\ref{eq:expectation_sigma}).
The discrepancy $D$ effectively measures the loss of the likelihood $\mathcal{L}_\mathrm{tot}$ of data $\{ \vec{\sigma}(t) \} _{t=0}^L$, \rev{as large $D$ corresponds to small $\mathcal{L}_\mathrm{tot}$ in general.}
%\noindent This function was first introduced in~\cite{Hoang2019Feb} to set a stopping criterion apart from the whole fitting process. 
\rev{Now we separately examine the model-data discrepancy for observed and missing data points by defining two measures:
\begin{align}
\label{eq:D2}
& D_{\mathrm{obs}} = \frac{1}{NL - |\mathcal{M}|} \sum_{(i,t+1) \notin \mathcal{M}} \big[\sigma_i (t+1) - \mathbb{E}[\sigma_i (t+1)] \big]^2, \nonumber \\
& D_{\mathrm{mis}} = \frac{1}{|\mathcal{M}|} \sum_{(i,t+1) \in \mathcal{M}} \big[ \sigma_i (t+1) - \mathbb{E}[\sigma_i (t+1)] \big]^2,
\end{align}
where $|\mathcal{M}|$ is the set size of missing data points. Note that we assume that initial data points do not include missing values as $(i, t=0) \notin \mathcal{M}$.}
%number of hidden data points belonging to the $i$-th spin.
%To separately examine the data-model consistency for observed and missing data points, we split $D$ as $D = D_{\mathrm{obs}} + D_{\mathrm{mis}}$, where $D_{\mathrm{obs}}$ and $D_{\mathrm{mis}}$ represent the discrepancy contribution from observed data points ($(i,t)\not\in \mathcal{M}$) and missing data points ($(i,t) \in \mathcal{M}$), respectively.
%\rev{How can these two quantities be used to find a good estimate of $W_{ij}$?} 
During the iterations of the EM steps, both  $D_{\mathrm{obs}}$ and $D_{\mathrm{mis}}$ keep decreasing because the M-step optimizes \rev{$\theta$} to minimize the \rev{model-data discrepancy}.
However, the relative size of $D_{\mathrm{obs}}$ and $D_{\mathrm{mis}}$ changes with iterations as follows.
%Therefore, the separate measures, $D_{\mathrm{obs}}$ and $D_{\mathrm{mis}},$ can provide a good stopping criterion for the EM iterations as follows:  
At the beginning, $D_{\mathrm{mis}}$ is larger than $D_{\mathrm{obs}}$ because the missing data points of \rev{$\sigma^m_i(t)$} are just randomly assigned. 
After some iterations, however, the missing data points can become overly fine-tuned whereas the observed data points are always fixed.
This makes $D_{\mathrm{mis}}$ smaller than $D_{\mathrm{obs}}$.
Therefore, it is intuitive to halt the iterations when the $D_{\mathrm{obs}}$ and $D_{\mathrm{mis}}$ curves cross each other to \rev{avoid overestimation of model parameters. The equal model-data consistency of $D_{\mathrm{obs}} = D_{\mathrm{mis}}$ implies that the stochastic model has the same uncertainty for predicting observed and missing data points. In other words, the restoration of missing data points becomes good enough for the model not to distinguish the observed and restored data points. We conduct various numerical experiments to demonstrate the power of this stopping criterion.}
%The crossing moment can be practically monitored by an inequality condition, $D_{\mathrm{mis}} - D_{\mathrm{obs}} < \epsilon$ with a small positive $\epsilon$.

%Then we observed that both functions gradually decrease with the iterations (see Figure~\ref{fig.first}, right panels). This is reasonable because $\sigma_i (t+1)$ moves toward its expected value, $\tanh{H_i (t)}$, throughout the iterations.

%Taking $D_{\mathrm{hid}}$ and $D_{\mathrm{obs}}$ into account separately helps us finding a good stopping criterion. During the first stage, $D_{\mathrm{hid}}$ would be greater than $D_{\mathrm{obs}}$ because hidden data have been just randomly assigned. The updated hidden data would be far different from their original values. In the long run, the hidden data would be excessively fine-tuned while the visible data remain the same. This makes $D_{\mathrm{hid}}$ getting smaller than $D_{\mathrm{obs}}$. Therefore, it is quite natural to halt the algorithm when the $D_{\mathrm{obs}}$ and $D_{\mathrm{hid}}$ cross each other to prevent overfitting. For practical reasons, especially when we deal with real data, we need a more relaxed condition; we stop the algorithm when $D_{\mathrm{hid}} - D_{\mathrm{obs}} < \delta$ with a small positive $\delta$.

Finally we summarize the overall procedure:

(\lowercase\expandafter{\romannumeral1}) Initialize missing data \rev{$\sigma^m_i (t)$} for $(i,t) \in \mathcal{M}$ with random binary values;

(\lowercase\expandafter{\romannumeral2}) M-step: Infer \rev{model parameters $\theta$} from the whole data using the logistic regression in Eq. (\ref{eq:logistic});

(\lowercase\expandafter{\romannumeral3}) E-step: Re-assign the missing data \rev{$\sigma^m_i (t)$} based on the likelihood ratio $\mathcal{L}_{i,t}^{\pm}/(\mathcal{L}_{i,t}^{+}+\mathcal{L}_{i,t}^{-})$ in Eq. (\ref{eq:update});

(\lowercase\expandafter{\romannumeral4}) Repeat (\lowercase\expandafter{\romannumeral2}) and (\lowercase\expandafter{\romannumeral3}) until $D_{\mathrm{mis}} - D_{\mathrm{obs}} < \epsilon$ holds with a small threshold $\epsilon$.

\rev{In this way, we can infer the model parameters, and restore missing data points.}
%obtain both the restored data and inferred coupling parameters.

%The novelty in our work compared to~\cite{Hoang2019Apr} lies in the stopping condition of the step (\lowercase\expandafter{\romannumeral4}).

\section{Results}
We test the performance of our method with simulated data using the kinetic Ising model.
After checking the concordance of inferred and true couplings with the simulated data, we apply the method to examine experimental recordings of neuronal activities. 

\label{results}
\subsection{Inference of kinetic Ising model}

\begin{figure}[t]
\centering
\includegraphics[width=8.6cm]{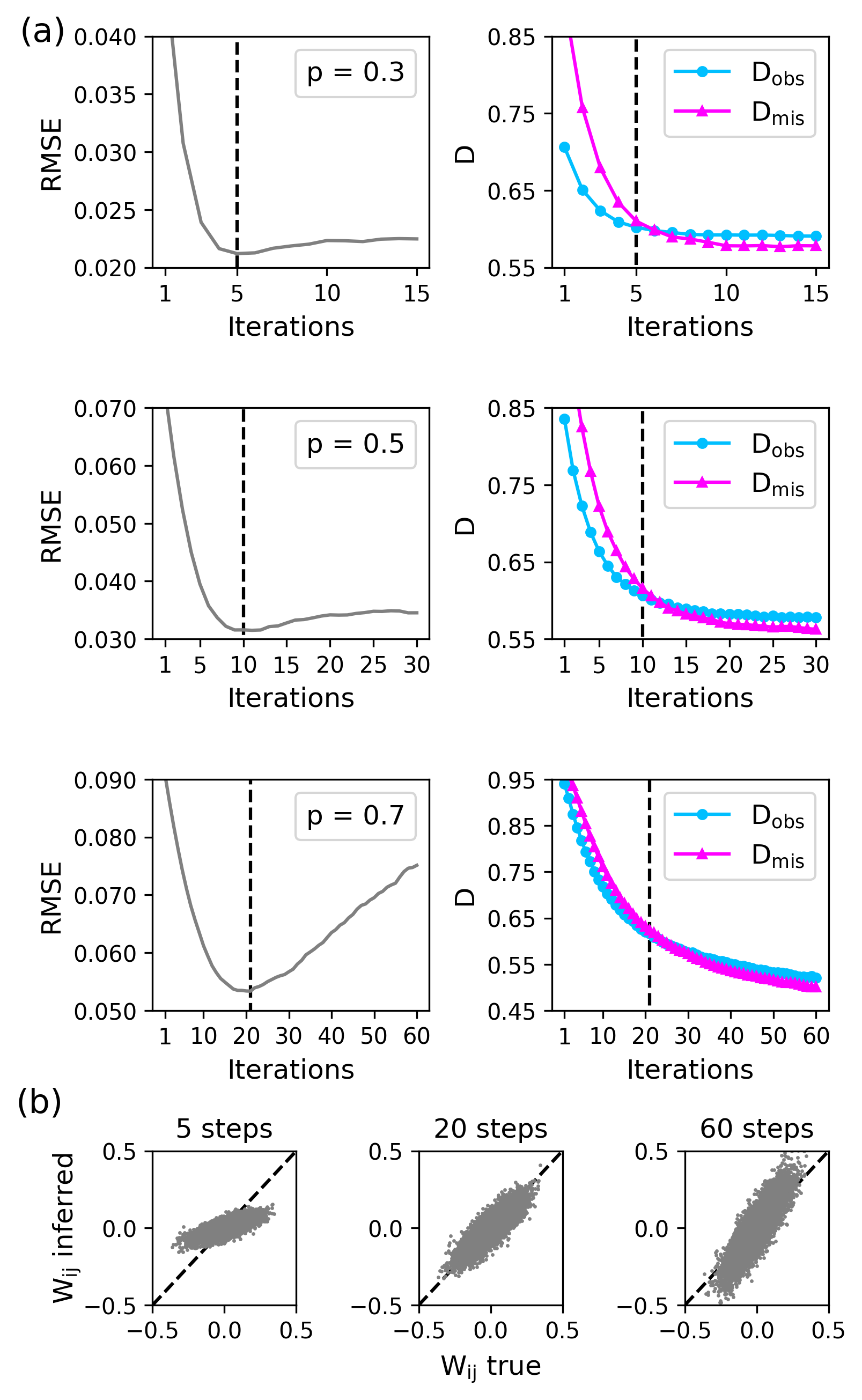}
\caption{\label{fig.first} \rev{(Color online) EM-based inference with missing data.} (a) The evolution of root-mean-square error (RMSE) of $W_{ij}$ (left panels) and two quality measures of inference, $D_{{\mathrm{obs}}}$ and $D_{{\mathrm{mis}}}$ (right panels), over the iteration of E- and M-steps. \rev{Dashed} lines denote the time when the algorithm stops according to our stopping criterion. The fraction of missing observations is $p = 0.3$, $0.5$, and $0.7$. (b) Comparison between the true and inferred $W_{ij}$ at different times: 5 (left), 20 (center), and 60 (right) steps after the iteration starts, with $p = 0.7$. The correct stopping point is near 20 steps according to our stopping criterion. A system size $N=100$ and a data length $L=10000$ are used.}
\end{figure}

\begin{figure}[t]
\centering
\includegraphics[width=8.6cm]{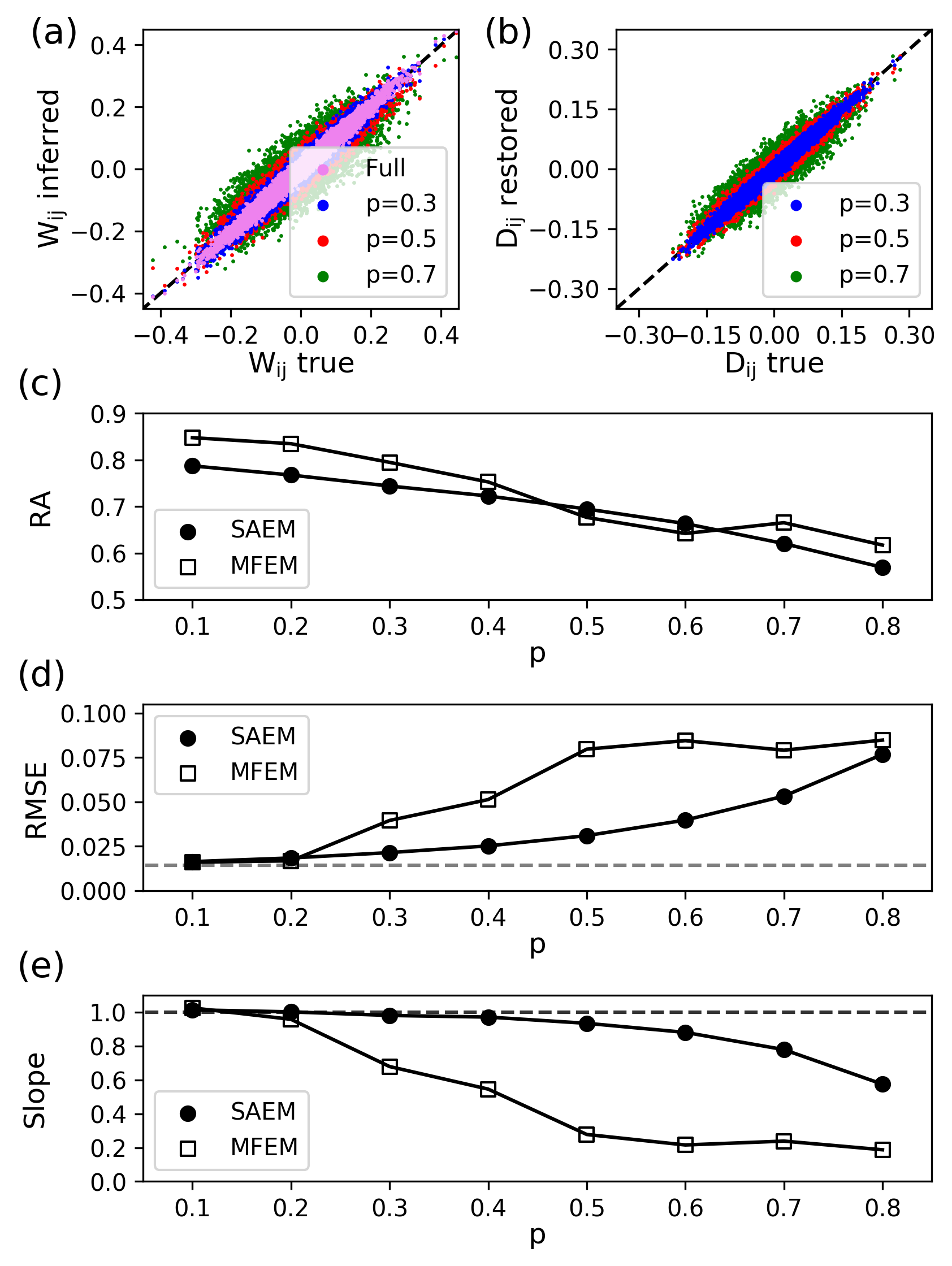}
\caption{\label{fig.second} \rev{(Color online) Quality of inference.} (a) Comparison between the true and inferred $W_{ij}$ with a fraction of missing observations $p = 0.3$ (blue), $0.5$ (red), and $0.7$ (green). A result from the full data (purple) is also shown. (b) One-step lagged correlation $D_{ij}$ obtained from true data and restored data. (c-e) Comparison of our stochastic approximation EM (SAEM, filled circles) and the mean-field EM (MFEM, empty squares) by Campajola \textit{et al.} \cite{Campajola2019Jun}, in terms of restoration accuracy (RA) of missing data (c), RMSE of $W_{ij}$ (d), and the slope of the linear regression between $W_{ij}$ and $W_{ij}^{\textrm{true}}$ (e).
The dashed lines in (d) and (e) corresponds to the RMSE and the slope inferred from the full data.
Here we show the average results of three independent experiments. The standard deviation is comparable to the marker size and thus not shown.
%\rev{RA and RMSE are compared between our stochastic EM method (filled circles) and a mean-field EM method (empty squares).}  
A system size $N=100$ and a data length $L=10000$ are used.
%{\color{red}Use black color only with filled circles and empty squares in (c), (d), and (e). It would be nice to show uncertainty of our inference by putting standard deviation of our inference in (c), (d), and (e). You may consider different single imputation. Change the legends as ``Our method'' and ``Mean-field'' into ``Stochastic EM'' and ``Mean-field EM''.}
}
\end{figure}

%To test the performance of our algorithm, we first used synthetic kinetic Ising data. 
We simulated a stochastic time series using the Sherrington-Kirkpatrick (SK) model~\cite{Sherrington1975Dec}.
The SK model follows the conditional probability in Eq.~(\ref{eq:prob}), where the local field is defined as $H_i = \sum_j W_{ij} \sigma_j + b_i$.
We turn off the bias $b_i = 0$ for simplicity, and consider asymmetric interactions $W_{ij} \neq W_{ji}$. 
In the SK model, each value of $W_{ij}$ is independently drawn from a Gaussian distribution, $\mathcal{N}(0, g^2/N)$, with zero mean and the variance of $g^2/N$.
We set the overall scale of the interaction as $g=1$.
With the selected $W_{ij}^{\mathrm{true}}$, we obtain $L=10000$ time points of $\vec{\sigma}(t) = (\sigma_1(t), \sigma_2(t), \cdots, \sigma_N(t))$ with a system size of $N=100$.
Then the total number of data values of $(i, t) \in \mathcal{A}$ has a set size \rev{$|\mathcal{A}| = (L+1)N$}.
\rev{We then mask some data points of $(i, t) \in \mathcal{M}$ as missing data points $\sigma^m_i(t)$, and hide the true $W_{ij}^{\mathrm{true}}$.}
The fraction of missing data is defined as $p = |\mathcal{M}|/|\mathcal{A}|$.
Our task is two-fold: to restore probable values of missing data points \rev{$\sigma^m_i(t)$} and to infer the true $W_{ij}^{\mathrm{true}}$. 
%with a system size $N=100$ and a data length $L=10000$. In this model, the elements of $J$ are independently drawn from the Gaussian distribution $\mathcal{N}(0,g^2/N)$. We set the scale of couplings $g=1$. Then we masked the data randomly and homogeneously, with a probability of $p$. 

%First, we validated our stopping criterion of the iterations (Figure~\ref{fig.first}).
As the EM algorithm iterates, we evaluated the quality of the inference by measuring the root-mean-square error of $W_{ij}$, $\mathrm{RMSE} = N^{-1} \sqrt{\sum_{i,j} (W_{ij}-W_{ij}^{\mathrm{true}})^2}$, as in~\cite{Campajola2019Jun}. 
Too many iterations usually lead to increasing RMSE (Figure~\ref{fig.first}(a), left panels).
In particular, the \rev{overshooting} becomes more evident when \rev{more data points are masked as missing data.}
At early iterations, $W_{ij}$ is underestimated because the missing data restored with random values reduce the correlation between variables (Figure~\ref{fig.first}(b), left panel). On the other hand, after too many iterations, $W_{ij}$ is overestimated because the discrete values of the restored missing data are even better fitted to the model, with exaggerated $H_i$ or larger values of $W_{ij}$ (right panel). An optimal iteration is where we find a reasonably unbiased estimation of $W_{ij}$ (center panel).
%We can also see clear evidence of overfitting by tracking the inferred $W_{ij}$ throughout the iterations (Figure~\ref{fig3}). Before the stopping point (left panel), $W_{ij}$ becomes underestimated because correlations between spins have been destroyed by randomly assigned missing data. After the stopping point (right panel), $W_{ij}$ becomes overestimated because the missing data have been too much adjusted. Around the stopping point we have found using $D_{\mathrm{obs}}$ and $D_{\mathrm{hid}}$ (center panel), we can observe unbiased estimation of $W_{ij}$.

Together with the RMSE, we examined the model-data \rev{discrepancy} measures of $D_{\mathrm{obs}}$ and $D_{\mathrm{mis}}$ for the observed and missing data points (Figure~\ref{fig.first}(a), right panels).
As expected, $D_{\mathrm{mis}}$ is larger than $D_{\mathrm{obs}}$ at the beginning of iterations when the restoration of missing data points is more or less random.
However, after some iterations, $D_{\mathrm{mis}}$ becomes smaller than $D_{\mathrm{obs}}$ since the restored data is excessively fine-tuned.
Interestingly, the equality of $D_{\mathrm{mis}}$ and $D_{\mathrm{obs}}$ occurs near the optimal iteration that minimizes the RMSE.
In practice, the crossing iteration can be monitored by an inequality condition, $D_{\mathrm{mis}} - D_{\mathrm{obs}} < \epsilon$ with a small positive $\epsilon$.
Here we take $\epsilon = 0.01$.

%\rev{This also explains the seemingly contradictory observation that the restoration accuracy (RA), defined by the ratio of correctly restored hidden data points, increases with excessive iterations despite the lower quality of the inference of $W_{ij}$ (Figure~\ref{fig.first}(c)). We again emphasize that imputation is different from prediction. The purpose of imputation is not to fill in the missing data as accurately as possible.}
%since smaller thresholds did not qualitatively affect the results.
%Figure~\ref{fig.first} shows that the minimum of RMSE corresponds to the intersection of $D_{\mathrm{obs}}$ and $D_{\mathrm{hid}}$, regardless of the value of $p$ (we used $\delta = 0.01$, but using smaller $\delta$ did not significantly affect the results). Especially when $p$ is large (bottom panels), our stopping criterion becomes crucial because the error diverges so quickly after a couple of iterations.

\begin{figure}[t]
\centering
\includegraphics[width=8.6cm]{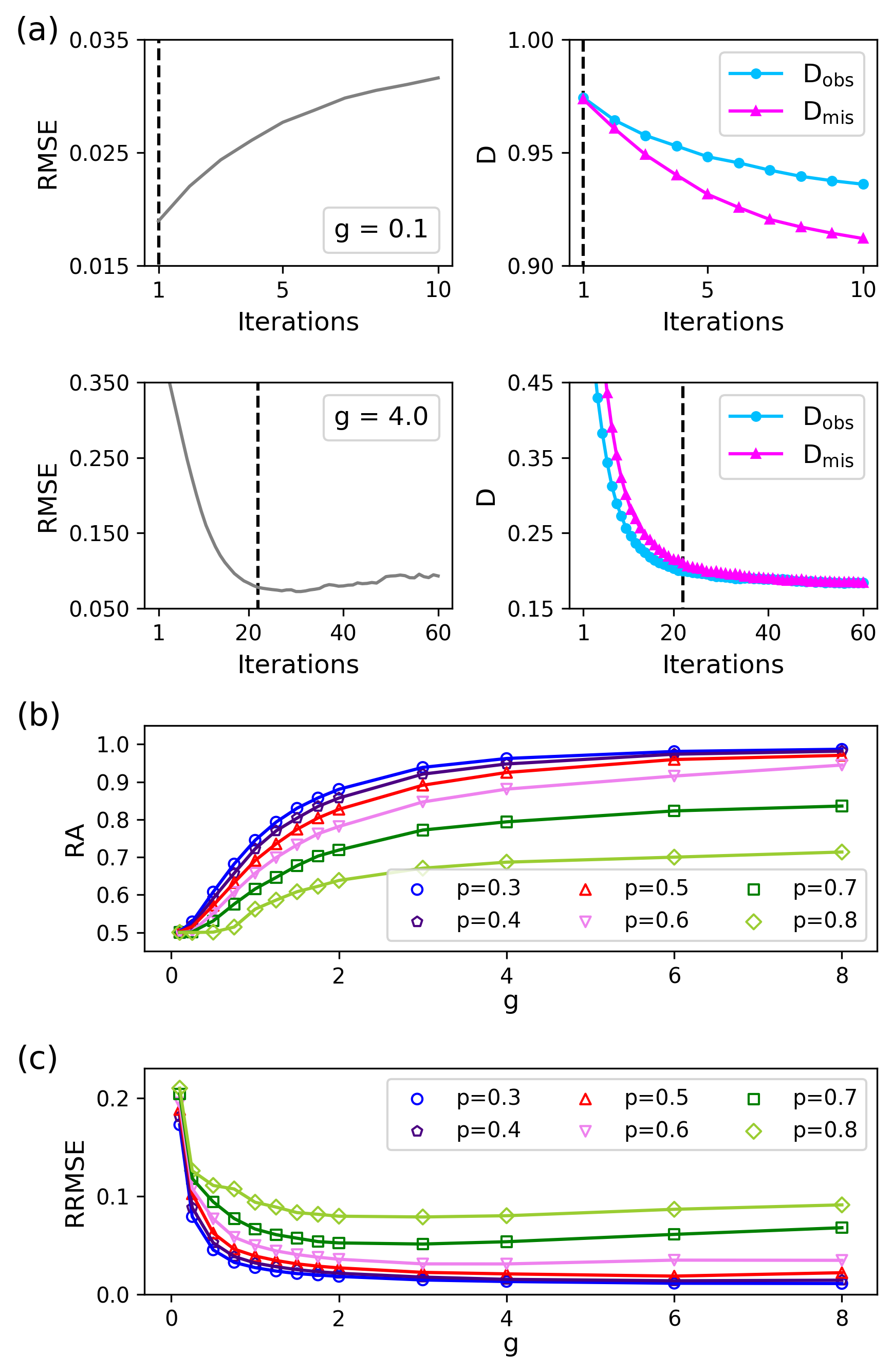}
\caption{\label{fig.g} \rev{(Color online) Inference for strong coupling. (a) The evolution of RMSE and $D$ over the iterations when the scale of couplings, $g$, is small ($g=0.1$, top panels) or large ($g=4.0$, bottom panels) with the fraction of missing observations $p=0.5$.} (b-c) RA of missing data (b) and rescaled RMSE of $W_{ij}$ (c) as a function of $g$ from $0.1$ to $8$. A system size $N=80$ and a data length $L=6400$ are used.}
\end{figure}

\begin{figure}[t]
\centering
\includegraphics[width=8.6cm]{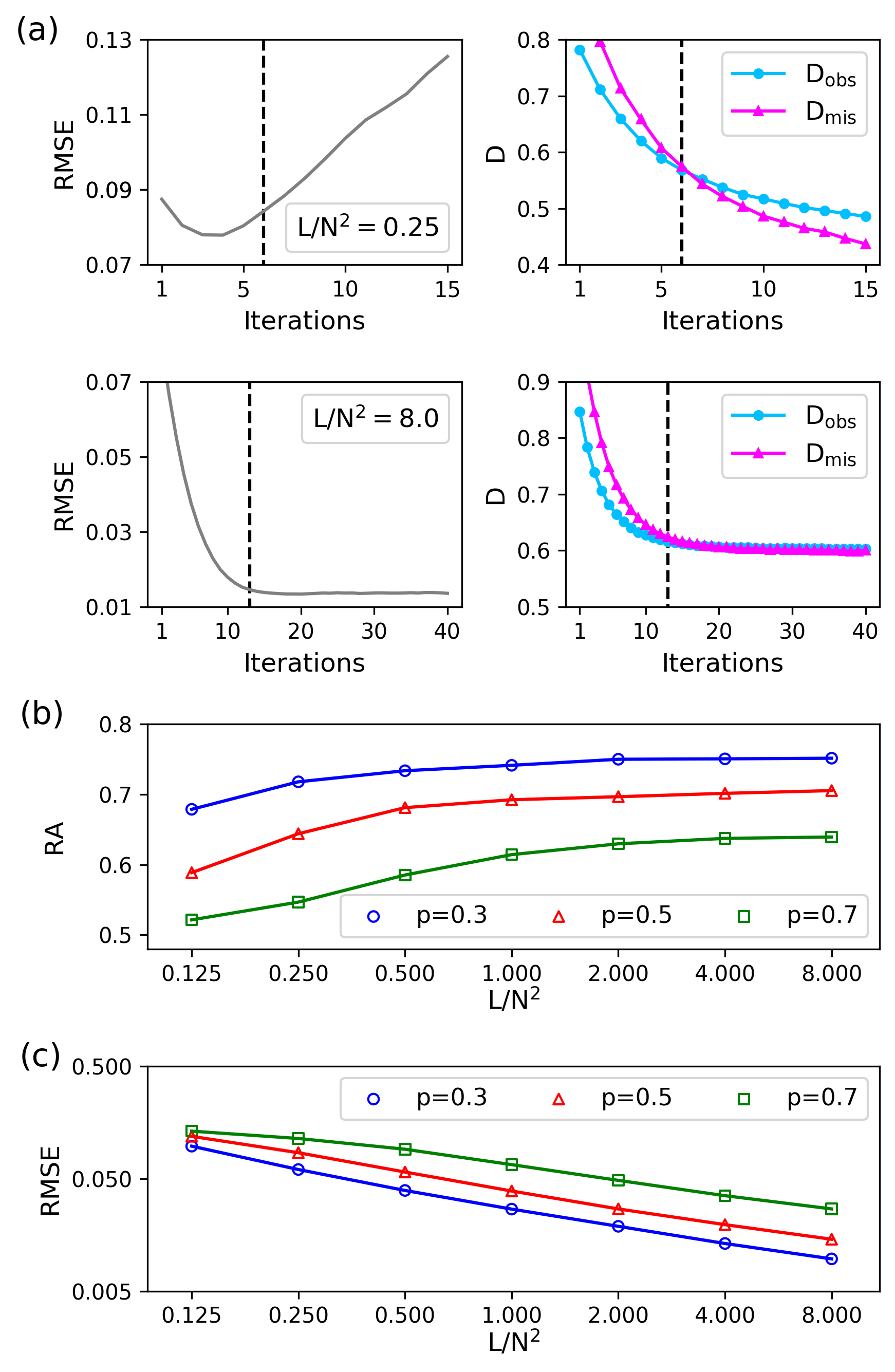}
\caption{\label{fig.L} \rev{(Color online) Inference and data size. (a) The evolution of RMSE and $D$ over the iterations when the data length, $L$, is short ($L/N^2=0.25$, top panels) or long ($L/N^2=8.0$, bottom panels)  with the fraction of missing observations $p=0.5$.} (b-c) RA of missing data (b) and RMSE of $W_{ij}$ (c) as a function of $L/N^2$ from 0.125 to 8. Here we use a system size of $N=80$, and the scale of couplings $g=1$.}
\end{figure}

After providing evidence for this stopping criterion, we checked the performance of our algorithm with varying fraction of missing data, $p.$
%Figure~\ref{fig.second}(a) is a scatter plot of the true versus inferred $W_{ij}$ for $p = 0$ (full data), $0.3$, $0.5$, and $0.7$. 
The result for $p=0.3$ is visibly indistinguishable from the one for full data \rev{corresponding to $p=0$}.
As expected, the inferred $W_{ij}$ deviates farther from its true value as $p$ increases (Figure~\ref{fig.second}(a)).
%This means that a missing fraction up to 30\% is bearable. 
Then we examined the collective behavior of data, particularly the one-step lagged correlation: $D_{ij} = \langle \sigma_i(t) \sigma_j(t+1) \rangle_t - \langle \sigma_i(t) \rangle_t \langle \sigma_j(t) \rangle_t $. Here \rev{$\langle f(t) \rangle_t \equiv 1/L \sum_{t=0}^{L-1} f(t)$} represents a time-averaged value of $f(t)$.
Note that $\langle \sigma_j(t+1)  \rangle_t \approx \langle \sigma_j(t) \rangle_t$ for large $L$.
The one-step lagged correlations are successfully recovered with our inference, even though a large fraction ($p = 0.7$) of data is missing (Figure~\ref{fig.second}(b)).
%We compared $D_{ij}$ computed from complete and restored data.
%Figure~\ref{fig.second}(b) shows that the one-step lagged correlations are successfully recovered with our inference even if a large fraction ($p = 0.7$) of data is missing.

%Then we calculated the one-step delayed correlations, $D_{ij} = \braket{\delta\sigma_i (t) \delta\sigma_j (t+1)}_t$, from both the true and restored data. Figure~\ref{fig.second}(b) shows that the algorithm successfully recovered correlations between spins even when a large fraction of data had been masked ($p = 0.7$).

\rev{We measured the accuracy of restoration and inference. First, restoration accuracy (RA) measures what fraction of the restored missing data $\sigma^m_i(t)$ is matched with the unmasked original data $\sigma_i(t)$.
Second, RMSE measures how well the inferred $W_{ij}$ is matched with true $W_{ij}^{\mathrm{true}}$.
In addition to the RMSE, we also measure the slope of the linear regression between $W_{ij}$ and $W_{ij}^{\mathrm{true}}$. As shown in Fig.~\ref{fig.first}(b), slope smaller than 1 represents underestimation of $W_{ij}$, whereas slope larger than 1 represents overestimation of $W_{ij}$.
Using these metrics, we compared our SAEM method to the mean-field EM (MFEM) method developed by Campajola \textit{et al.}~\cite{Campajola2019Jun} (see Appendix~\ref{appendix_campajola} for a brief summary of the method).}
%data and model inference can be measured by the restoration accuracy (RA) of missing data, RMSE, and slope of the linear regression, $a$ of $W_{ij}=aW_{ij}^{\mathrm{true}}+b$. We compared these metrics to the mean-field EM approach developed by Campajola \textit{et al.} \cite{Campajola2019Jun} (see Appendix~\ref{appendix_campajola} for the brief summary of the method). 

\rev{RA is the ratio of correctly restored missing data points, and it generally decreases with $p$ (Figure~\ref{fig.second}(c)). 
Our method restores nearly 80\% of the masked data points when 10\% of the full data is masked as missing data. However, if 80\% of the data is missing, the restoration is more or less the same as a random restoration.}
\rev{While MFEM achieves a slightly better RA than our SAEM, ours shows clearly better model inference with smaller RMSE especially in the intermediate range of $0.3 \leq p \leq 0.7$ (Figure~\ref{fig.second}(d)). Furthermore, it ensures an unbiased estimation of $W_{ij}$ at least when the $p$ is less than $0.5$ (Figure~\ref{fig.second}(e)). In contrast, MFEM suffers from severe underestimation of $W_{ij}$ when more than 20\% of the data is missing.}

\rev{Now we test the performance of SAEM in harsher inference conditions with too weak or strong couplings or an insufficient data size.}
Throughout the experiments, we used a fixed system size $N=80.$ Unless mentioned otherwise, the data length is $L=6400$ and the scale of couplings is $g=1.0$.

\rev{First, we evaluated the effect of coupling scale $g$. Strong spin interactions lower stochasticity in the kinetic Ising model. When $g$ is too small ($\approx 0.1$), every missing data point has a flip probability close to 50\%, so accurate restoration of missing data is fundamentally impossible. This is why $D_\mathrm{mis}$ becomes smaller than $D_\mathrm{obs}$ right after random initialization (Figure~\ref{fig.g}(a), upper-right panel). At this point, we can also achieve the lowest error of $W_{ij}$, as the RMSE keeps increasing over the iterations (upper-left panel). However, large $g$ ($\approx 4.0$) makes $D_\mathrm{mis}$ and $D_\mathrm{obs}$ converge more slowly (bottom panels). Therefore, the exact value of $\epsilon$, the threshold value of $D_\mathrm{mis} - D_\mathrm{obs}$, affects the stopping point of the EM iterations more significantly, although RMSE does not change much with iterations for the converging case.
A larger $g$ results in a higher RA (Figure~\ref{fig.g}(b)). This is expected because strong $g$ makes the inference of missing data less stochastic. For a large $g > 4$ and small $p < 0.5,$ almost all missing data are correctly restored. For a small $g < 0.25$, however, RA becomes around 50\%. 
Next we examined the quality of model inference depending on $g$.
To ensure a fair comparison between results with different scales of coupling, we used a rescaled RMSE, $\mathrm{RRMSE} = \mathrm{RMSE}/g,$ as used in~\cite{Campajola2019Jun} (Figure~\ref{fig.g}(c)). For $p < 0.5,$ RRMSE keeps decreasing with $g$. Similar to the worse restoration, this is because small $g$ induces larger fluctuations with a flip probability close to 50\%, making the inference more difficult. When $p > 0.6$, the RRMSE curves deviate from those of $p < 0.5$, especially when $g$ is large. Taken together, these results indicate that our SAEM is applicable to a wide range of $0.5 < g <  8$, at least when less than 50\% of data is missing.}

\rev{Next, we checked the dependency of our SAEM on the data length $L.$ When $L$ is small ($L/N^2 = 0.25$), RMSE increases quickly after the optimal iteration, which emphasizes an important role of our stopping criterion (Figure~\ref{fig.L}(a), top panels). When $L$ is large ($L/N^2 = 8$), however, both RMSE and data-model discrepancy $D$ converge (bottom panels). 
As more data is provided, RA increases and finally saturates (Figure~\ref{fig.L}(b)), and RMSE of $W_{ij}$ decreases (Figure~\ref{fig.L}(c)). It is of interest that RMSE shows a power-law-like behavior in the wide range of $L/N^2$, which was also observed in~\cite{Campajola2019Jun}. The power-law exponent does not depend on $p$. Therefore, this scaling behaviour can suggest an appropriate sample size for the applicability of our SAEM.}

\rev{To further validate SAEM and our stopping criterion, we surveyed its performance under various conditions. We investigated the effects of sparsity or symmetry of coupling $W_{ij}$ and the presence of external bias $b_i$. 
Then we confirmed that this method shows accurate restoration and inference under these conditions (refer Appendix~\ref{appendix_experiment} for details).}

\subsection{Inference of neuronal dynamics}
\label{real data}

\begin{figure*}[t]
\centering
\includegraphics[width=18cm]{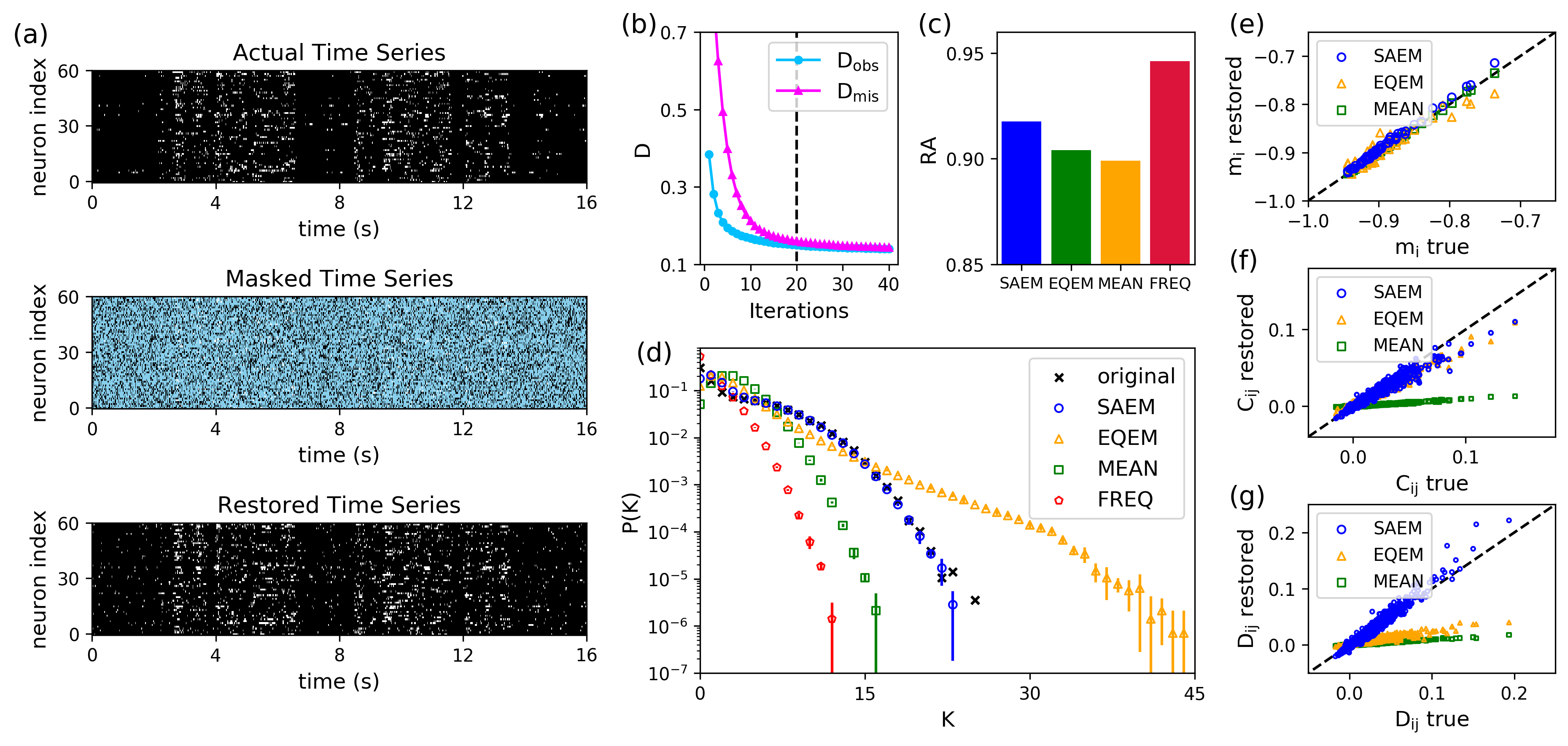}
\caption{\label{fig.neuron1} \rev{(Color online) Inference with real neural activity data.} (a) Restoration of neuronal activity data published by Marre \textit{et al.}~\cite{neuron1}. Among neural activities of 60 neurons (top), we randomly mask a part of them with a fraction of $p = 0.7$ (center). Then we recover the masked data using our stochastic EM-based inference method (bottom). \rev{(b) The evolution of model-data discrepancy $D$ for observed and missing data over the EM iterations.} (c) Restoration accuracy (RA) of missing data points is compared between four methods: our non-equilibrium EM model (SAEM), equilibrium EM model (EQEM), \rev{imputation by mean activities of observed neurons (MEAN), and imputation by the most dominant activity (FREQ)}. Five independent experiments are averaged. (d) Relative frequencies of $K$ simultaneous spikes obtained from the original unmasked data and restored data from four inference methods. (e-g) Comparison between true and restored mean activity $m_i$ (e), equal-time activity correlation $C_{ij}$ (f) one-step lagged activity correlation $D_{ij}$ (g) for three inference methods except for FREQ.
%\color{red}{Please change NEQ $\rightarrow$ SAEM and EQ $\rightarrow$ EQEM to be consistent with other figures.}
}
\end{figure*}

%\begin{figure}[t]
%\centering
%\includegraphics[width=8.6cm]{fig.neuron1}
%\caption{\label{fig.neuron1} \rev{(a) Test of our algorithm on neuronal activity data published by Marre \textit{et al.}~\cite{neuron1}. Among the neural activities of 60 neurons (top), we randomly mask them with a probability of $p = 0.7$ (center). Then we recover the data using our inference algorithm (bottom). (b) The evolution of $D$ over the iterations. (c) RA of the four data restoration algorithms based on a nonequilibrium model (NEQ), equilibrium model (EQ), imputation by the mean activities (MEAN), and imputation by the most frequent value (FREQ). Five independent experiments are averaged.}}
%\end{figure}

%\begin{figure}[t]
%\centering
%\includegraphics[width=8.6cm]{fig.neuron2}
%\caption{\label{fig.neuron2} \rev{(a) Measured and inferred probability of $K$ simultaneous spikes. (b-d) Comparison between true and restored mean activity $m_i$ (d), one-step lagged correlation $D_{ij}$ (e), and equal-time correlation $C_{ij}$ (f) for the the three inference algorithms besides FREQ.}}
%\end{figure}

Having confirmed that our algorithm robustly works for synthetic data, we now apply the algorithm to explore real data.
%After validating our algorithm with synthetic data, we wondered whether it also works with experimental data. Therefore, 
We examined temporal data of neuronal activities recorded in salamander retinal ganglion cells~\cite{neuron1}. %focusing on its data reconstruction ability. 
The data consist of neuronal spike trains of 160 neurons, whose bin size is 20 ms. The state of each bin is considered {\it active} ($\sigma_i (t) = +1$) when at least one spike is present in the time bin, and {\it inactive} ($\sigma_i (t) = -1$) otherwise. \rev{We assumed that the effect of measurement noise is negligible in the binarized data.} The total length of the time series is $L = 283041$.
In this study, we \rev{randomly selected 60 neurons among the 80 most active neurons} to exclude silent neurons. %that show low firing rates for most of the time.
Figure~\ref{fig.neuron1}(a) summarizes our problem setting. We have neuronal activity recording data (top panel). After masking \rev{70\%} of them randomly as missing data (center panel), we restore the missing data using our inference method (bottom panel). %We examined three different data restoration schemes.

\rev{First, we monitored the evolution of $D_{\mathrm{obs}}$ and $D_{\mathrm{mis}}$ during the iterations of the SAEM method (Figure~\ref{fig.neuron1}(b)). The gap between two model-data discrepancies decreases with the iterations. However, $D_{\mathrm{mis}}$ never falls below $D_{\mathrm{obs}}$. We have also observed that the crossover between $D_{\mathrm{mis}}$ and $D_{\mathrm{obs}}$ diminishes as data size gets larger (Figure~\ref{fig.L}(a), bottom panels). The experimental data has a pretty large data size ($L/N^2 \approx 80$). In this case, the specific value of $\epsilon$ becomes crucial to determine the stopping point of the EM iteration. Here we used $\epsilon=0.01$ as in the previous application for the synthetic data because it is good enough to accurately restore missing data and reproduce statistical features in data.}

To evaluate the inference performance, we compare \rev{four inference methods: SAEM, EQEM, MEAN, and FREQ}.
SAEM and EQEM are based on physical models.
SAEM implements the kinetic or non-equilibrium Ising model. We have elaborated on this algorithm in previous sections.
EQEM utilizes the standard equilibrium Ising model, which has been widely used to model collective behavior of neuronal networks~\cite{Shlens2006Aug,Schneidman2006Apr,Tkacik2014Jan}. This algorithm uses an EM-based approach similar to SAEM, but it considers the neural activities at different times separately (see Appendix~\ref{appendix_eq} for the details of EQEM).
For both algorithms, we include external bias $b_i$ as well as neighboring interactions $W_{ij}$.
\rev{Contrary to SAEM and EQEM, MEAN and FREQ use simple imputation schemes. MEAN stochastically assigns $\sigma^m_i(t) = \pm 1$ under a constraint that their mean activity becomes equal to the mean activity of observed data points $\sigma^o_i(t)$. FREQ simply assigns $\sigma^m_i (t) = -1$ for every missing data point considering that silent neurons are dominant in the observed data.}
%{\color{red}Please check the above sentences for MEAN and FREQ.}

\rev{First, we compared RA of the four methods (Figure~\ref{fig.neuron1}(c)).
Unexpectedly, FREQ, the simplest restoration method, achieves the highest RA (94.6\%). Other methods show lower accuracy: 91.8\% (SAEM), 90.4\% (EQEM), and 89.9\% (MEAN). Despite  this fact, the fundamental problem of FREQ is that it cannot consider the inherent stochasticity of missing data points. Although this naive method performs the best restoration in terms of RA, it overlooks a small portion of data points with $\sigma^m_i (t) = +1$, and thus perturbs all the collective properties present in the data.} 
%NEQ shows the highest restoration accuracy (RA = 91.1\%), compared with 88.1\% for NULL and 89.6\% for EQ (Fig.~\ref{fig7}(b)). 
%However, the restoration accuracy of NEQ is not outstanding.
%This is because the neurons are mostly inactive (see Fig.~\ref{fig7}(a), top panel).
%Indeed, if we simply assign all the missing data as \textit{inactive} ($\sigma_i(t) = -1$), this restoration shows a better RA = 93.5\% than the three methods even including NEQ.
%Even though $Noneq$ achieved the highest RA and $Indep$ yielded the lowest, the three methods showed basically the similar RA. This would be because the neurons were most of the times inactive (see Figure~\ref{fig7}(a), top panel). Indeed, if we simply assume that all of the missing observations are $-1$, the RA becomes 93.5\% which is higher than any other methods. 
%However, this naive restoration would perturb all the collective properties in data. Therefore, RA is not an appropriate estimator of the quality of data restoration.

Therefore, we focused on other descriptive statistics of data related to the collective behavior of neurons, to compare the inference performance.
First, we measured the number of simultaneous spikes, $K(t) = \sum_{i=1}^N \big(\sigma_i(t) + 1 \big)/2$, at different times, and obtained their relative frequencies $P(K)$ (Figure~\ref{fig.neuron1}(d)).
We found that SAEM matches the original $P(K)$ much better than any other methods. \rev{MEAN and FREQ} underestimate a probability of large $K$, while EQEM predicts a much heavier tail for $P(K)$ than the original one. This tendency for EQEM was also observed by Tka\v{c}ik \textit{et al.}~\cite{Tkacik2014Jan}. They tried to solve the issue by using an equilibrium model with pairwise interactions and an additional potential parameterized by $K.$ 
Here we emphasize that SAEM, having non-equilibrium dynamics into the system, naturally captures the pattern of simultaneous firing without further assumptions.
\rev{The marked underestimation of $P(K)$ by FREQ is expected because it considers all missing data points as inactive ($\sigma^m_i (t) = -1$).}
%Here we present an alternative approach: implementing nonequilibrium dynamics into the system. Then we can catch the observed patterns of simultaneous firing without further assumptions.

Next, we examined mean activities $m_i = \braket{\sigma_i(t)}_t$, equal-time correlations $C_{ij} = \braket{\sigma_i (t) \sigma_j (t)}_t - \braket{\sigma_i (t)}_t\braket{\sigma_j (t)}_t$, and one-step lagged correlations $D_{ij} = \braket{\sigma_i (t) \sigma_j (t+1)}_t - \braket{\sigma_i (t)}_t\braket{\sigma_j (t)}_t$ from the restored data (Figure~\ref{fig.neuron1}(e-g)).
%Next, we compared the mean activities and pair correlations among the neurons, calculated from the three reconstructed data. From Figure~\ref{fig7}(d)-(f) we can see that 
SAEM correctly infers all the statistics, $m_i$, $C_{ij},$ and $D_{ij}$.
The correct estimation of $C_{ij}$ is surprising because SAEM matches $m_i$ and $D_{ij}$ by tuning $b_i$ and $W_{ij}$ and does not directly affect $C_{ij}.$ 
Similarly, the correct estimation of $m_i$ and $C_{ij}$ by EQEM is not surprising because EQEM models $m_i$ and $C_{ij}$ by tuning $b_i$ and $W_{ij}.$
However, EQEM severely underestimates $D_{ij},$ perhaps because neuronal firing is far from an equilibrium process in that it is a manifestation of information transmission.
Finally, \rev{MEAN} by definition fits only $m_i$ and obviously fails to account for pairwise correlations between spins, both $C_{ij}$ and $D_{ij}.$ \rev{We omit the results for FREQ, since it is obvious that FREQ cannot reproduce all of the statistics.}
%is only an indirectly inferred observable. Contrastingly, \textit{Eq} infers $m_i$ and $C_{ij}$ successfully but severely underestimates $D_{ij}$ which is not concerned by an equilibrium model. In addition, \textit{Eq} matches $m_i$ less accurately than $Noneq$. 
%{\color{blue}Considering that the number of the fitting parameters for \textit{Noneq} and \textit{Eq} is nearly the same ($N+N(N+1)/2$ for \textit{Noneq} and $N+N(N-1)/2$ for \textit{Eq}), we can conclude that \textit{Noneq} has a better ability to restore neural spike train data than \textit{Eq}. }
%Lastly, $Indep$ matches $m_i$ correctly but fails to fit $D_{ij}$ and $C_{ij}$ since it completely ignores correlations. 

As a further corroboration of the validity of SAEM, we examined another data set of neuronal activities~\cite{neuron2}, and confirmed that SAEM reproduces the collective behavior of neurons as shown here (see Appendix~\ref{appendix_neuron}).
These results clearly demonstrate that our method (SAEM) is an effective approach for understanding real experimental data.

\section{Discussion}
\label{discussion}

We developed a stochastic approximation EM algorithm that infers the kinetic Ising model from stochastic time series with missing data.
The algorithm alternates between an E-step \rev{stochastically} restoring missing data points and an M-step optimizing model parameters.
Using this \rev{SAEM with an appropriate stopping criterion for the EM iterations}, we could successfully infer model parameters without under- or over-estimation even when up to 70\% of the data is missing. 
We demonstrated the performance of the inference with synthetic data from extensive simulations of the kinetic Ising model, and with real neuronal data. In particular, our algorithm, based on a non-equilibrium model, outperforms equilibrium models in reproducing collective behavior in neuronal activities.

We found an effective scheme for determining the optimal number of iterations that provides the best model inference. 
\rev{Under easy inference conditions such as a moderate coupling regime, sufficient data, and a small fraction of missing data, excessive EM iterations do no harm since the inferred parameter values ultimately converges.
However, under difficult inference conditions, the excessive iterations lead to worse inference of coupling strengths in the kinetic Ising model.
Here, for the best inference, it is crucial to stop SAEM at the right iteration.
Our key idea is that the stochastic model must show equal uncertainty to predict observed and missing data points because seamless restoration of missing parts implies no distinction between observed and restored parts. This condition can be practically monitored by the equality of the model-data discrepancy of observed and missing data ($D_{\mathrm{obs}}=D_{\mathrm{mis}}$).}
\revi{We confirmed that the stopping criterion works well in various conditions. However, it still lacks theoretical justification or derivation from fundamental physical principles.}

This study applied the \rev{SAEM} algorithm for the kinetic Ising model.
However, other non-equilibrium models can be considered as the underlying stochastic model. 
%Other out-of-equilibrium models than the kinetic Ising model can also be utilized. 
For instance, Marre \textit{et al.}~\cite{Marre2009Apr} proposed an Ising model incorporating both spatial and temporal correlations among neurons, and showed that this model predicts spatio-temporal patterns of neuronal activities significantly better than the standard Ising model (also see~\cite{Nasser2013Mar,Nasser2014Apr}). 
%Moreover, Also, we can add some kind of nonstationary behavior to the original kinetic Ising model. 
Moreover, Tyrcha \textit{et al.}~\cite{Tyrcha2013Mar} showed that applying non-stationary external bias can explain a structure of neuronal spike trains even without direct interactions between neurons. %Incorporating these concepts into our approach would be another direction for future research.

Our methodology to deal with incomplete kinetic Ising data can be applied to other fundamental scenarios. For example, we can apply our algorithm to data with irregular observation times. Unequally spaced time series data is common in astronomy~\cite{Lomb1976Feb,Scargle1982Dec}, paleoclimatology~\cite{Schulz2002Apr}, biology~\cite{Ruf1999Apr}, and many other fields in which regular observations are infeasible. In the context of the kinetic Ising model, sequences at unobserved times can be correctly inferred using exactly the same approach as this paper. Other applications would be the reconstruction of interaction networks when experimental artifacts make sporadic missing observations. We expect that our new algorithm can be generalized for many situations.

\section*{Acknowledgment}
\rev{We thank Carlo Campajola for kindly sharing the code for the mean-field EM method.} This work was supported by the Intramural Research Program of the National Institutes of Health, NIDDK (V.P.), and the New Faculty Startup Fund from Seoul National University, and the National Research Foundation of Korea (NRF) grant funded by the Korea government (MSIT) (Grant No. 2019R1F1A1052916) (J.J.).

\setcounter{figure}{0}
\renewcommand{\thefigure}{A\arabic{figure}}
\appendix
\section{\rev{Mean-field EM}}
\label{appendix_campajola}

\rev{We briefly summarize the mean-field EM (MFEM) developed by Campajola \textit{et al.} \cite{Campajola2019Jun}.
The first step of MFEM is to initialize the coupling parameters, $W_{ij}$, with random values. An analytic approximation of expected log-likelihood, with the current coupling parameters and observed data points, can be derived using the Martin-Siggia-Rose path integral formalism \cite{Martin1973Jul} and the saddle-point approximation. This gives a self-consistent equation for calculating the mean magnetization, $m_i (t)$, for each missing data point. After solving this equation to compute $m_i (t)$ (E-step), the gradient of the expected log-likelihood with regard to $W_{ij}$ is obtained. Applying the gradient ascent gives the update rule for $W_{ij}$ (M-step). Repeating these two steps until convergence, one can get both the coupling parameters $W_{ij}$ and the missing data points $\sigma^m_i (t) = \mathrm{sign}(m_i (t))$.
The authors also observed an underestimation of $W_{ij}$ (see Figure~\ref{fig.second}(e)). To mitigate this issue, they proposed a recursive procedure that iterates the whole EM algorithm several times. Every time the algorithm ends, they fix the most polarized $m_i (t)$ for each time $t$ according to its sign. This can reduce the extent of underestimation, and noticeably improve the quality of inference \cite{Campajola2019Jun}. However, this procedure requires multiple implementations of the original algorithm, requiring longer computation time.}

\begin{figure}[t]
\centering
\includegraphics[width=8.6cm]{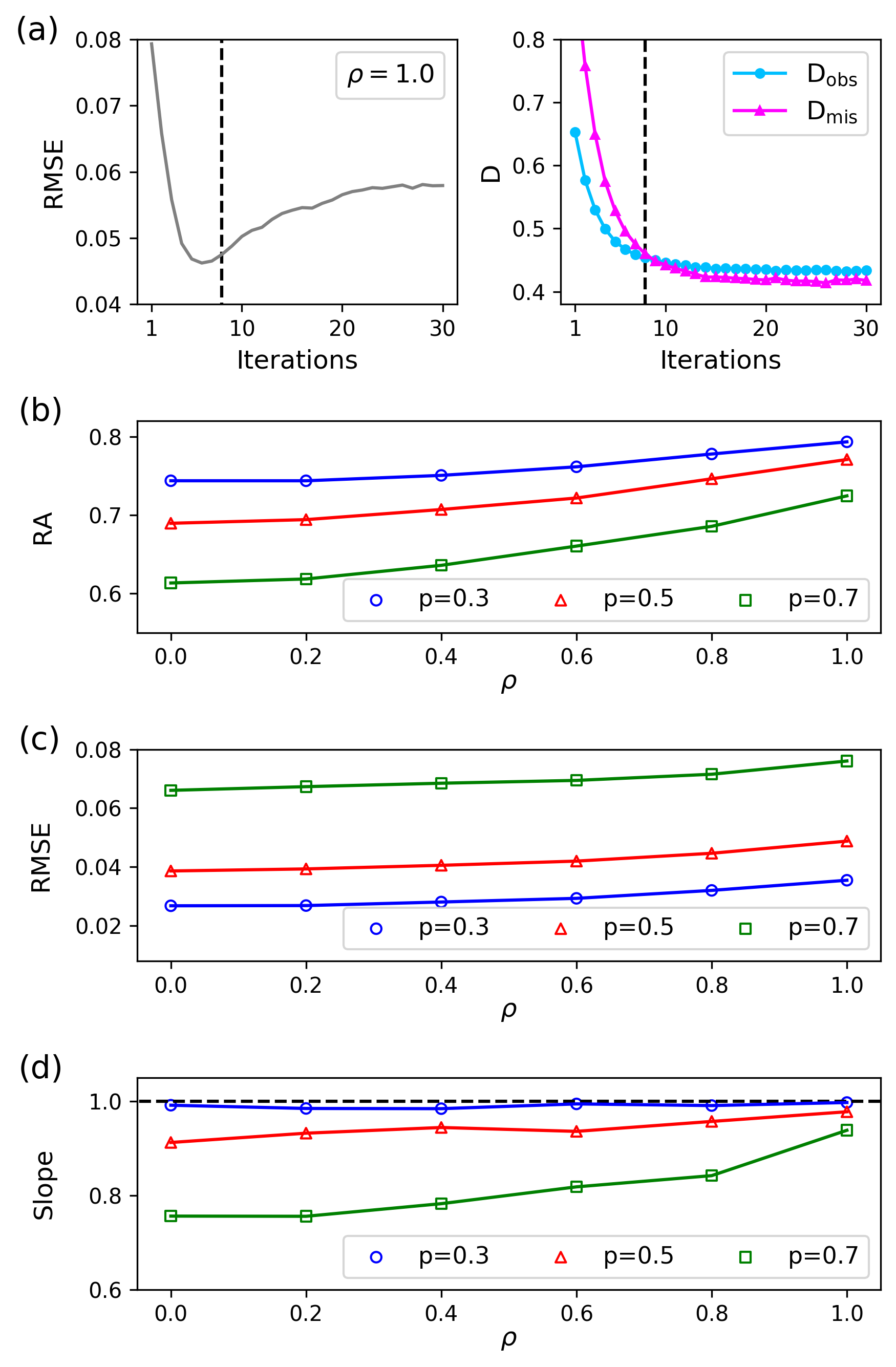}
\caption{\label{fig.rho} \rev{Inference of symmetric coupling. (a) The evolution of RMSE and $D$ over the iterations when a coupling matrix $W$ is fully symmetric (symmetry parameter $\rho=1$) and a fraction of missing data is $p=0.5$. (b-d) RA of missing data (b), RMSE of $W_{ij}$ (c), and the linear regression slope (d) as a function of $\rho$.}}
\end{figure}

\section{\rev{Additional validations}}
\label{appendix_experiment}

\rev{We further tested the robustness of our SAEM under various conditions such as sparsity or symmetry of coupling strengths, and the presence of external biases.}

\begin{figure}[t]
\centering
\includegraphics[width=8.6cm]{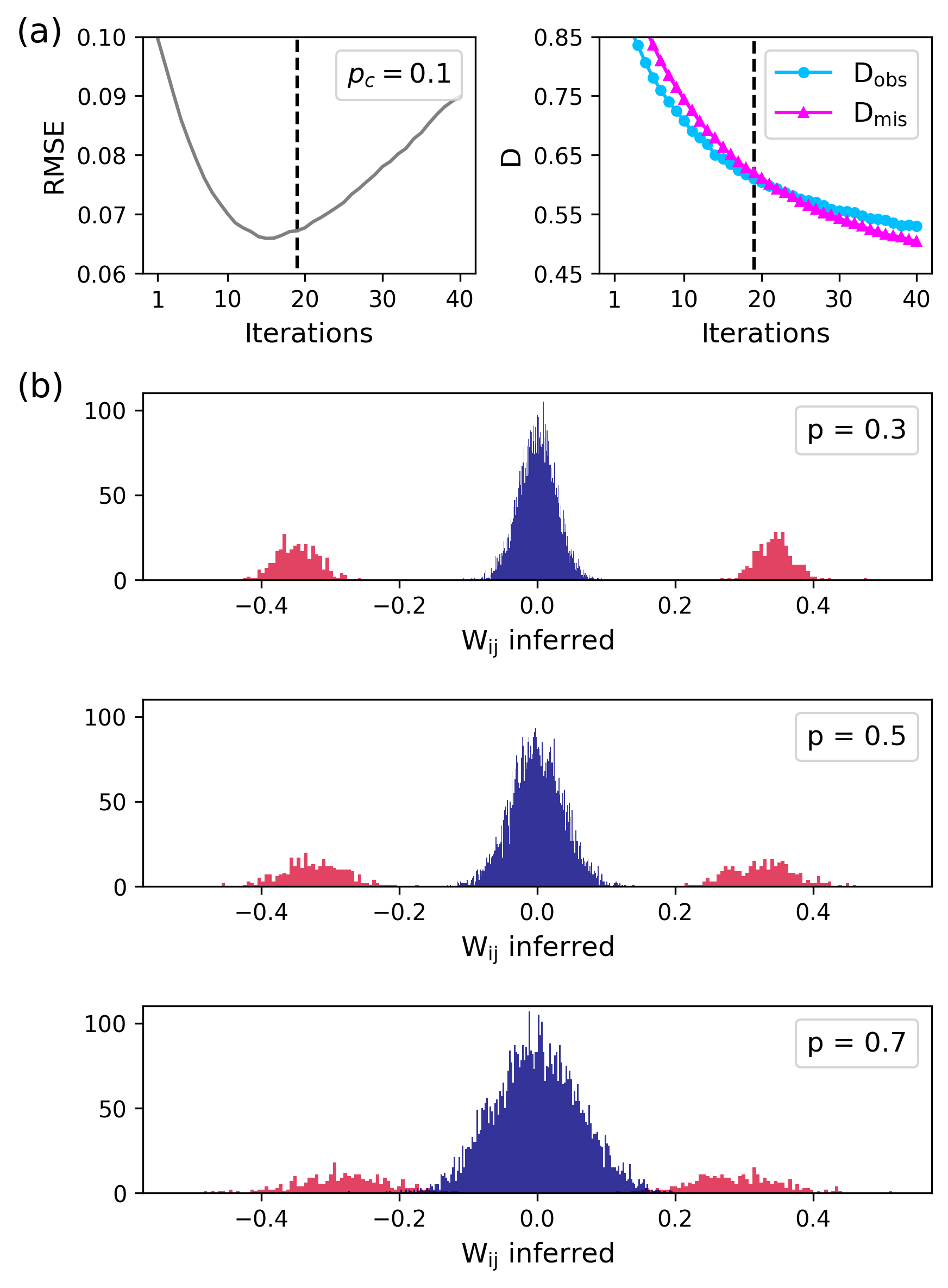}
\caption{\label{fig.pc} \rev{Inference of sparse coupling. (a) The evolution of RMSE and $D$ over the iterations when interactions are sparse (sparsity $p_c = 0.1$). A fraction of missing data is $p=0.7.$ (b) Inference of connected (red) and disconnected (blue) couplings with $p=0.3$ (top), $0.5$ (center), and $0.7$ (bottom).}}
\end{figure}

\begin{figure}[t]
\centering
\includegraphics[width=8.6cm]{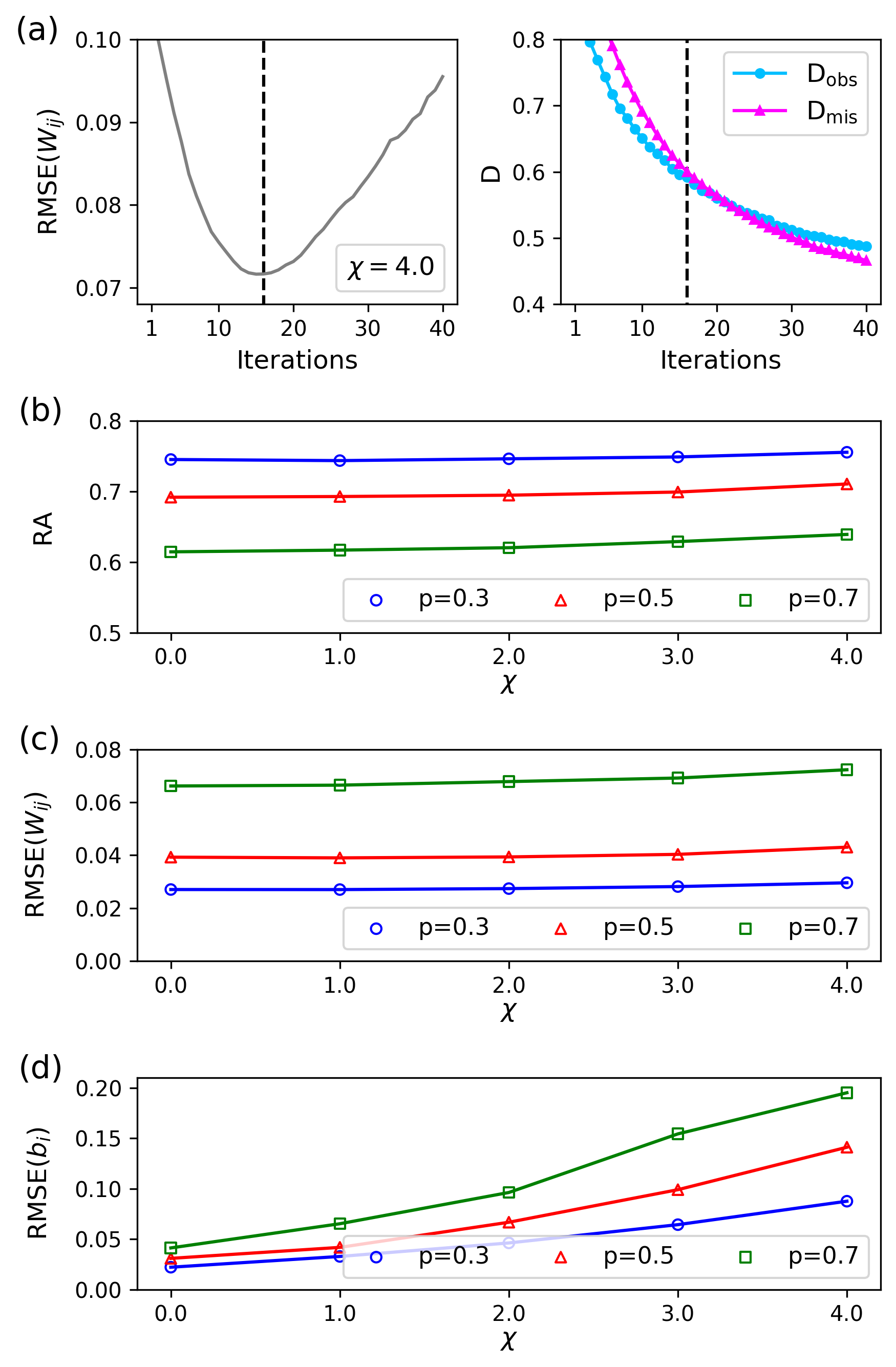}
\caption{\label{fig.bias} \rev{Inference of strong bias. (a) The evolution of RMSE (of $W_{ij}$) and $D$ over the iterations when there is a constant bias ($b_i = \chi g/\sqrt{N},$ where $\chi=4.0$) for spins. A fraction of missing data is $p=0.7.$ (b-d) RA of missing data (b), RMSE of $W_{ij}$ (c) and $b_i$ (d) as a function of $\chi$.}}
\end{figure}

%The inference performance usually depends greatly on the interaction strength $g$ and the length $L$ of time series. Thus we examine their dependency for the restoration and inference by varying $g$ and $L$ with a fixed system size $N = 80.$
%First, we found that RA increases as $g$ increases (Fig.~\ref{fig.g}(a)).
%We also tested our algorithm with varying $g$ and $L$. We used $N=80$ in these tests. Figure~\ref{fig.g}(a) shows that the RA increases with the scale of couplings $g$.

\rev{First, we investigated the effect of symmetry of coupling strengths. The original SK model \cite{Sherrington1975Dec} assumes that all couplings are independent and identically distributed. We mitigated this assumption by inducing symmetry to the coupling matrix: $W=\sqrt{1-\rho} W_{\mathrm{rand}} + \sqrt{\rho} W_{\mathrm{sym}}$. The elements of $W_{\mathrm{rand}}$ and $W_{\mathrm{sym}}$ are drawn from a Gaussian distribution, $\mathcal{N}(0, g^2/N),$ but $W_{\mathrm{sym}}$ is symmetric. $\rho \in [0,1]$ is a control parameter tuning the degree of symmetry. For example, $\rho=1$ makes the coupling matrix fully symmetric.
We confirmed that the stopping criterion ($D_{\mathrm{obs}}-D_{\mathrm{mis}}<\epsilon$) worked to find an optimal iteration when a coupling matrix is fully symmetric (Figure~\ref{fig.rho}(a)). It is of interest that RA slightly increases as $W$ becomes more symmetric (Figure~\ref{fig.rho}(b)), while RMSE gets slightly worse (Figure~\ref{fig.rho}(c)). We also measured the linear regression slope of $W_{ij}^{\mathrm{true}}$ on $W_{ij}$ and found that it increases with $\rho$ (Figure~\ref{fig.rho}(d)). Given a small fraction of missing data ($p < 0.5$), the slope is always close to $1.0,$ implying no under- or overestimation.}

\rev{Second, we examined the effect of sparsity of coupling strengths. We consider a discrete SK model where the couplings are $W_{ij} \in \{-g/\sqrt{p_c N}, 0, g/\sqrt{p_c N}\}$ with a probability of $p_c/2$, $1-p_c$, and $p_c/2$, respectively, following \cite{Aurell2012Mar}. %{\color{red}{Please check this sentence. I am not sure how $p_c$ changes the sparsity of coupling strengths. I expect we should control the density of zero-amplitude coupling, not the amplitude of coupling strength.}} 
Here $p_c$ is a control parameter tuning the  sparsity of a network, \textit{i.e.}, a fraction of connected pairs.
For a sparse network with $p_c=0.1,$ we applied our algorithm to infer missing time series. Our stopping criterion again found a near-optimal point (Figure~\ref{fig.pc}(a)), even when a large amount of data is masked ($p=0.7$). The inferred couplings are clearly separated (Figure~\ref{fig.pc}(b)). We can easily distinguish connected pairs from disconnected ones regardless of a fraction of missing data $p$. Note that we did not use other techniques such as $\ell_1$-regularization~\cite{Ravikumar2010Jun,Lokhov2018Mar} or decimation of couplings~\cite{Decelle2014Feb,Decelle2015May}.}
%Implementing these techniques to our scheme would be for future research.}

\rev{Lastly, we activated the bias $b_i$. When $b_i$ was drawn from a Gaussian distribution with zero mean, our algorithm found a good stopping point as well as reasonable estimates of $W_{ij}$ and $b_i$. We did not show the results because they are not significantly different from the case when $b_i=0.$
How about when all $b_i$ have the same sign, either plus or minus? For instance, neurons are usually inactive, so $b_i$ for a neuronal network would be mostly negative. To deal with this problem, we set a fixed $b_i = \chi g/\sqrt{N}$, where $\chi$ denotes the scale of a bias. Our stopping criterion worked well with a moderately high bias ($\chi=4.0,$ Figure~\ref{fig.bias}(a)). RA and RMSE of $W_{ij}$ do not significantly change with $\chi$ (Figure~\ref{fig.bias}(b-c)). Meanwhile, RMSE of $b_{i}$ increases with $\chi$ as expected (Figure~\ref{fig.bias}(d)).}

\section{Equilibrium Ising Model}
\label{appendix_eq}

\begin{figure}[t]
\centering
\includegraphics[width=8.6cm]{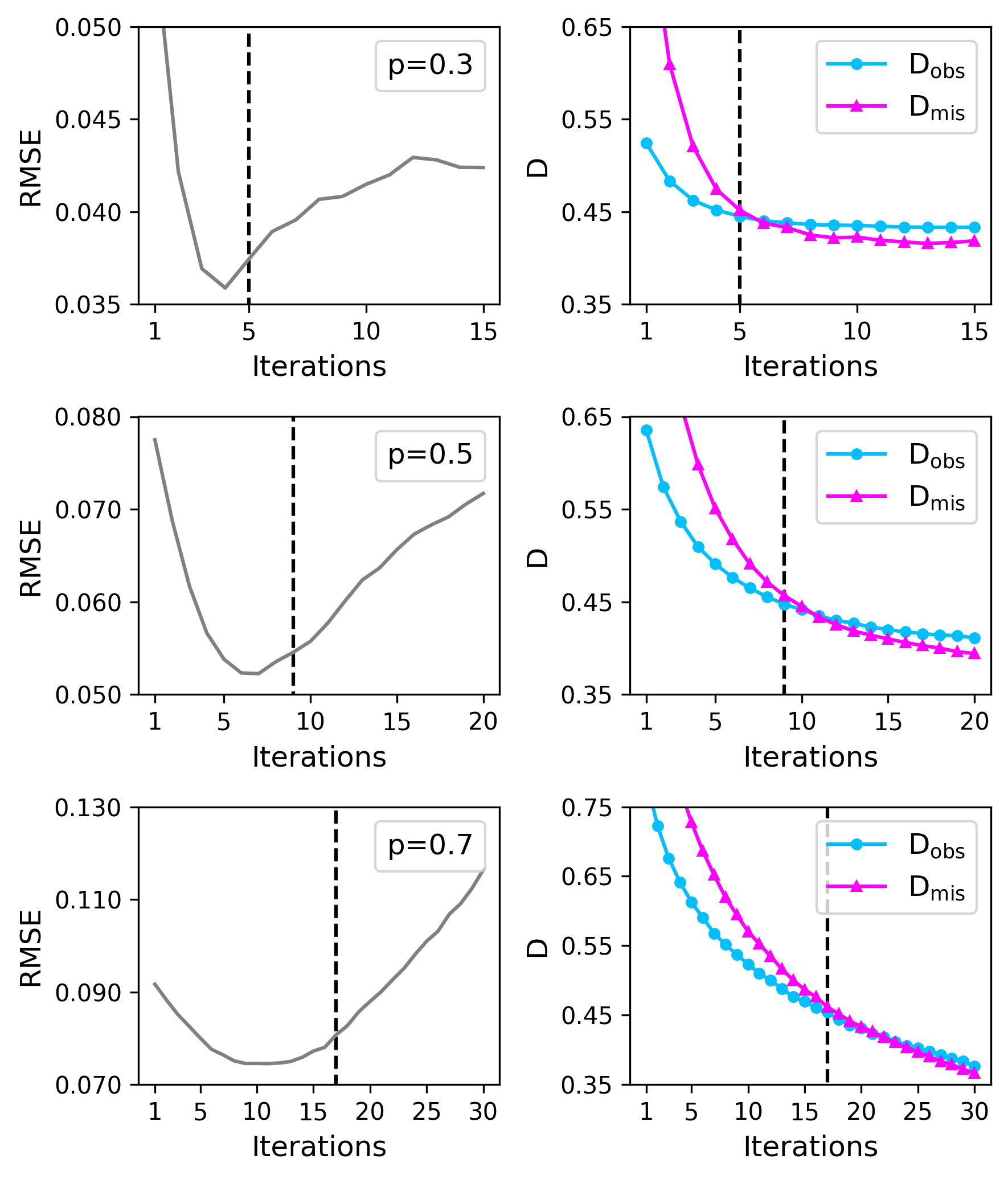}
\caption{\label{fig8} The same as Figure~\ref{fig.first}(a) but with the equilibrium Ising model.
}
\end{figure}

\begin{figure}[t]
\centering
\includegraphics[width=8.6cm]{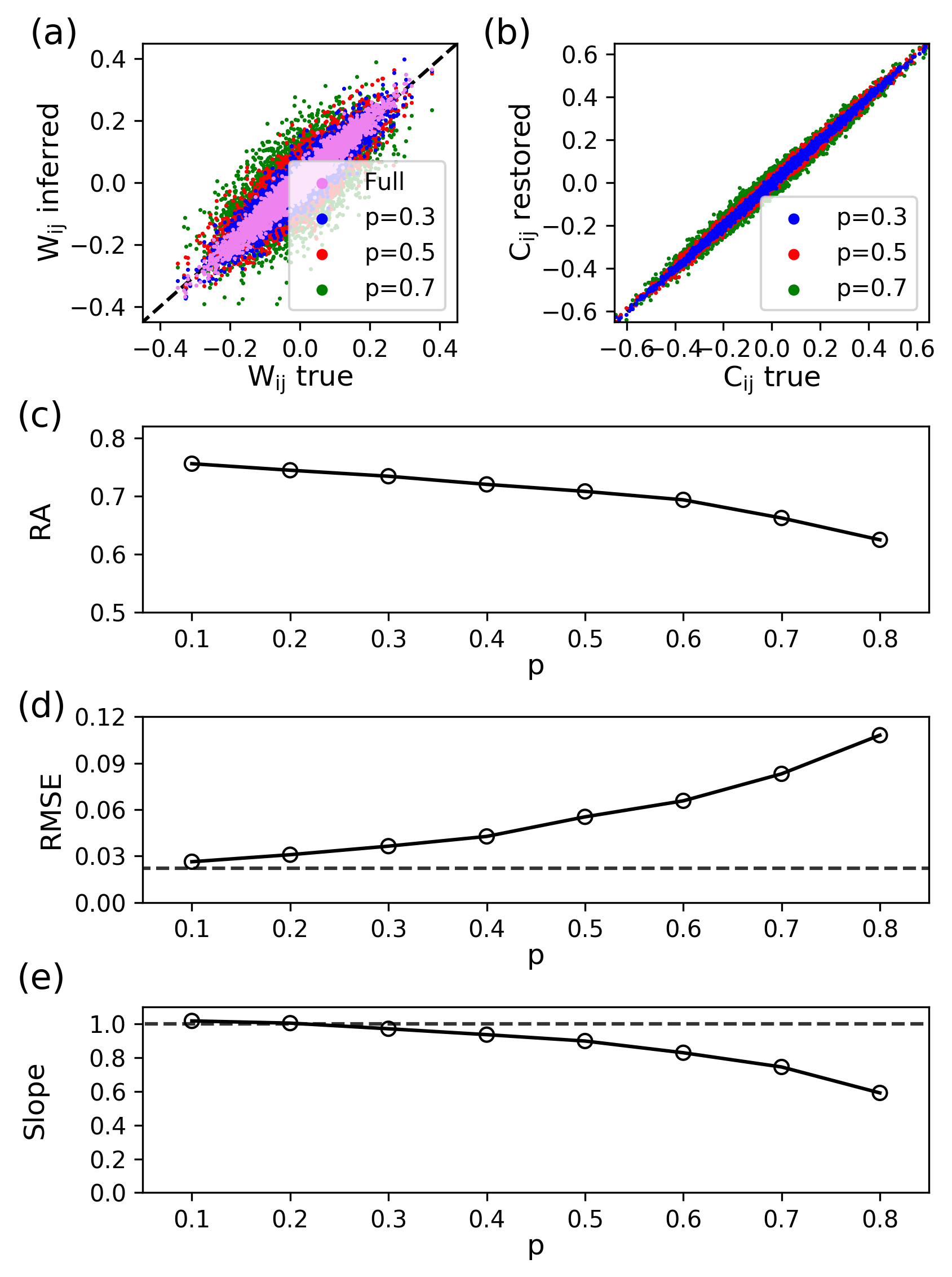}
\caption{\label{fig9} The same as Fig.~\ref{fig.second} but with the equilibrium Ising model. The one-step lagged correlation $D_{ij}$ in Figure~\ref{fig.second}(b) is replaced by the equal-time correlation $C_{ij}$.
}
\end{figure}

\begin{figure*}[t]
\centering
\includegraphics[width=18cm]{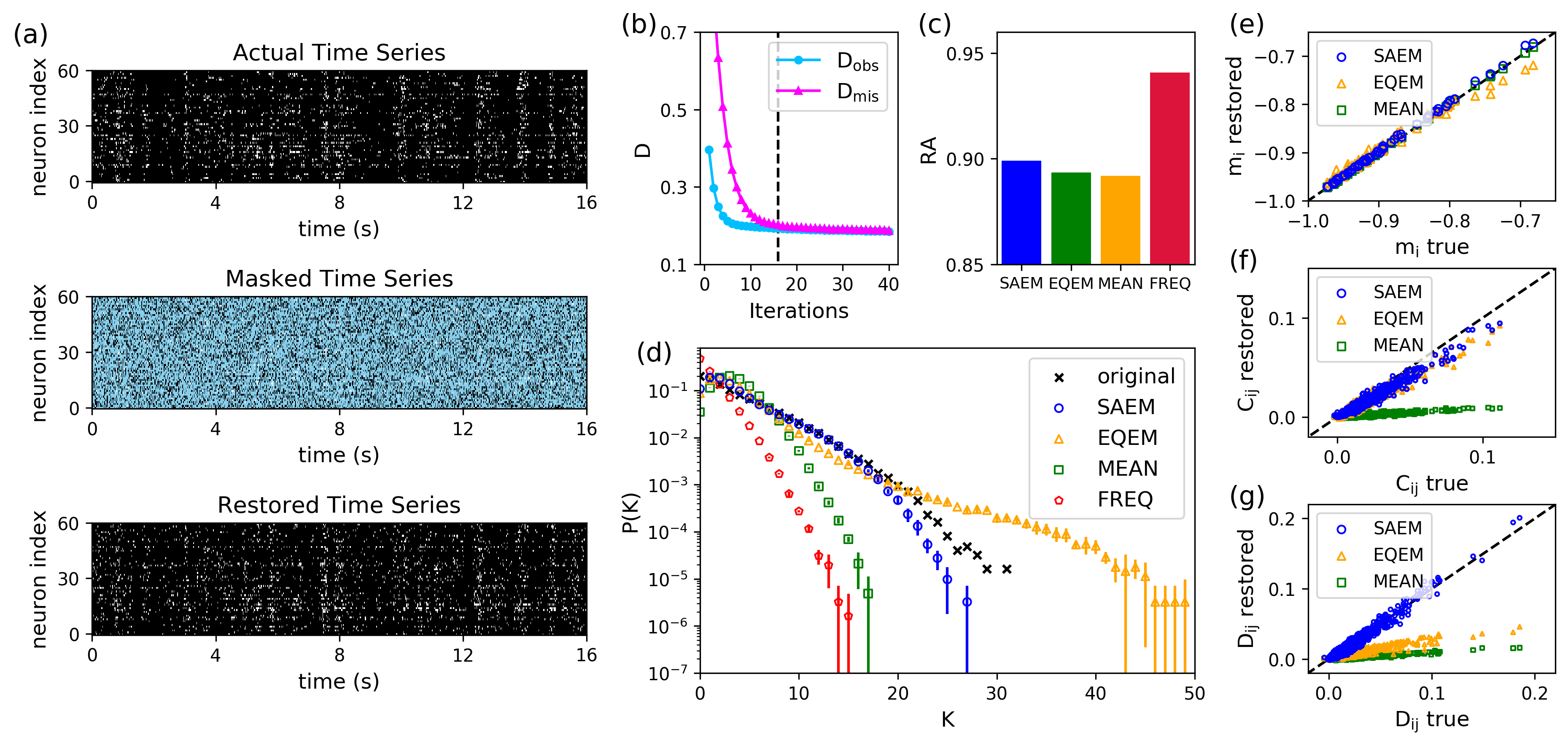}
\caption{\label{fig.neuron3} The same as Figure~\ref{fig.neuron1} but with the data published by Kohn and Smith~\cite{neuron2}.
}
\end{figure*}

%\begin{figure}[t]
%\centering
%\includegraphics[width=8.6cm]{fig.neuron4}
%\caption{\label{fig.neuron4} The same as Figure~\ref{fig.neuron2} but with the data published by Kohn and Smith~\cite{neuron2}
%}
%\end{figure}

In this section, we describe the \rev{equilibrium EM (EQEM)} method based on the equilibrium Ising model, which we have briefly mentioned in Section~\ref{real data}. In the standard Ising model, a probability of spin configuration $\vec{\sigma} = (\sigma_1, \sigma_2, ..., \sigma_N)$ with $\sigma_i = \pm 1$ is parameterized by an energy $E(\vec{\sigma}),$ $P(\vec{\sigma}) = Z^{-1} \exp{[-E(\vec{\sigma})]}$. Here, $Z = \sum_{\vec{\sigma}} \exp{[-E(\vec{\sigma})]}$ is a normalization factor and called a partition function. Energy is described by the first and second moments of spins, \textit{i.e.}, $E(\vec{\sigma}) = -\sum_{i < j} W_{ij} \sigma_i \sigma_j - \sum_i b_i \sigma_i$, where $W_{ij}$ is an interaction parameter and $b_i$ an external bias.
%(again we neglect an external field parameter $h_i$). 
The task is to reconstruct missing data points and find the true \rev{$\theta = (W_{ij}, b_i)$} when some data points are missing.

We again split the algorithm into E- and M-steps. First, we initialize missing data points with random binary values. Note that in the equilibrium model, the time index $t$ merely acts as a label to distinguish different sequences of $\vec{\sigma}$. From the randomly assigned data, we infer an error-prone $\theta$. Then we restore the missing data points with $\pm 1$ using the ratio of two probabilities, \rev{$P(F_i^{+}(\vec{\sigma}(t))|\theta) / P(F_i^{-}(\vec{\sigma}(t))|\theta)$} (E-step). $F_i^{\pm}(\vec{\sigma}(t))$ is defined similar to the main text (see Section~\ref{method}).
\rev{Calculation of the ratio of probabilities does not require the computation of partition function and therefore can be easily done.}
We update all of the missing data points one by one. %We checked that increasing the number of updates in one step did not affect the results.

Given the restored data points, we can infer the optimal \rev{$\theta$} (M-step) by maximizing the data likelihood, called the Boltzmann machine~\cite{Ackley1985Jan}. However, this requires an extensive calculation of partition function $Z,$ which entails the sum of $2^N$ configurations. 
Physicists have developed a number of methods to circumvent the exact computation of $Z$, including mean-field~\cite{Kappen98boltzmannmachine,Tanaka1998Aug}, Bethe approximation~\cite{Ricci-Tersenghi2012Aug}, Sessak-Monasson~\cite{Sessak2009Jan}, and adaptive cluster expansions~\cite{Barton2016Oct} (see also a pedagogical review~\cite{Nguyen2017Jul}). 
In this study, we choose a maximum pseudolikelihood approach~\cite{Aurell2012Mar} that maximizes the {\it local} pseudolikelihood, \rev{$\mathcal{L}_i(\theta) = \prod_t P(\sigma_i(t)|\vec{\sigma}_{i}^c (t),\theta)$ for each $i$, where $\vec{\sigma}_{i}^c$ denotes all spins besides the $i$-th spin. Since $P(\sigma_i(t)|\vec{\sigma}_{i}^c (t),\theta)$} is determined by a local field $H_i (t) = \sum_{j \ne i} W_{ij} \sigma_j (t) + b_i$, the same logistic regression function~\cite{scikit-learn} can be utilized, allowing fast and accurate inference. After optimizing each \rev{$\mathcal{L}_i(\theta)$} separately, we put $W_{ij} \leftarrow (W_{ij}+W_{ji})/2$ to ensure symmetric couplings.

Running this EQEM, we observed the overshooting similar to what we have observed from the algorithm with kinetic Ising model (SAEM, see Figure~\ref{fig.first}). To avoid overshooting, we define the same model-data discrepancy measures for observed and missing data points, $D_\mathrm{obs}$ and $D_\mathrm{mis}$, in which $\sigma_i (t+1)$ is replaced to $\sigma_i (t)$ (see Eq.~(\ref{eq:D2})). We track the evolution of $D_\mathrm{obs}$ and $D_\mathrm{mis}$ and halt the EM iteration when the stopping condition $D_\mathrm{mis} - D_\mathrm{obs} < \epsilon$ is met (we use $\epsilon = 0.01$ as in the main text). Overall, EQEM and SAEM are nearly the same, but the main differences are (i) we apply a new stochastic rule to restore missing data points and (ii) we use a pseudolikelihood maximization scheme to infer $W_{ij}.$

Figure~\ref{fig8} illustrates our stopping criterion for EQEM. Samples were generated using the Metropolis-Hastings algorithm \cite{Hastings1970Apr} with a system size $N = 100$ and a sample size $L = 10000.$ 
Especially when a fraction of missing data is large ($p=0.7$), the iteration stops several steps after we achieve the minimum error of $W_{ij}$, which means that the model inference of EQEM is suboptimal.
In contrast, SAEM showed coincidence of the minimum of RMSE and the intersection of $D_\mathrm{obs}$ and $D_\mathrm{mis}$ in general cases. 
\rev{The mismatch in EQEM may result from the pseudolikelihood approximation. EQEM used the pseudolikelihood for the inference (M-step), whereas it used the exact likelihood for the restoration (E-step).}

Even with this seemingly improper stopping criterion, EQEM is capable of inferring the true $W_{ij}$ (Figure~\ref{fig9}(a)). Furthermore, an excellent agreement between the pair correlations $C_{ij}$ of true and restored data was found (Figure~\ref{fig9}(b)), despite a relatively large deviation of the inferred $W_{ij}.$
Next, we addressed the quality of data and model inference with restoration accuracy of missing data, root-mean-square error of $W_{ij}$, and linear regression slope of true $W_{ij}$ on inferred $W_{ij}$. RA is above 0.6 even when 80\% of the data is missing (Figure~\ref{fig9}(c)). RMSE decreases with $p$ as expected (Figure~\ref{fig9}(d)). When a small fraction of data is missing ($p < 0.4$), we can accurately infer $W_{ij}$ with an error less than 0.05. In this regime, the slope of linear regression is close to 1.0, which means no under- or overestimation (Figure~\ref{fig9}(e)). These results are similar to SAEM (Figure~\ref{fig.second}).

\section{Other experimental data}
\label{appendix_neuron}

In addition to Section~\ref{real data}, we tested our algorithms with other neuronal spike train data published by Kohn and Smith~\cite{neuron2}. They recorded spiking activities of 70-100 neurons in the visual cortex of adult monkeys under various visual stimuli. Here, we chose a ``spontaneous'' dataset where a uniform gray screen had been used. We binned the neural activity with 20 ms and selected 60 neurons randomly from the 80 most active neurons. The data length was $L = 123404$. For the detailed experimental procedure, see~\cite{Smith2008Nov,Kelly2010Dec}. 
We summarize the results in Figure~\ref{fig.neuron3} with the same format as Figure~\ref{fig.neuron1}. The results are similar.

\bibliographystyle{apsrev4-1}
\bibliography{missing}

\end{document}